\documentclass{CSML}

\def\dOi{11(3:25)2015}
\lmcsheading%
{\dOi}
{1--31}
{}
{}
{Jan.~13, 2014}
{Sep.~30, 2015}
{}

\ACMCCS{[{\bf Theory of computation}]: Models of computation}

\usepackage{verbatim}
\usepackage{latexsym}
\usepackage{hyperref}
\usepackage{url}
\usepackage{amsmath}
\usepackage{amssymb}
\usepackage{mathrsfs}
\usepackage{stmaryrd}
\usepackage{bm}
\usepackage{color}
\usepackage{longtable}
\usepackage[T1]{fontenc}
\usepackage{listings}

\lstdefinelanguage{scheme}{ 
language=lisp, morekeywords={and,backquote,begin,cond,define,else,if,lambda,let,let*,letrec,or,quasiquote,quote,set!,unquote,unquote-splicing},
sensitive=false,
morecomment=[l]{;},
morestring=[b]",
  breaklines=true,
 breakatwhitespace=true,
  showstringspaces=true, 
 upquote=true,
 basicstyle=\ttfamily
}

\newcommand{\Yoleg}{Y^{*}}
\newcommand{\YolegVar}{Y_{\mbox{\scriptsize{Curried}}}^{*}}
\newcommand{\uline}[1]{\underline{#1}}
\newcommand{\uuline}[1]{\underline{\underline{#1}}}
\newcommand{\Maps}[3]{{#1}\stackrel{#2}{\Longrightarrow}{#3}}
\def\agengen/{\agen/ generalization}
\def\ldef/{$\lambda$-definability}
\def\labs/{$\lambda$-abstraction}
\def\lisp/{LISP/Scheme}
\def\boehm/{B{\"o}hm}
\def\BA/{Bracket-Abstraction}
\def\ba/{bracket-abstraction}
\newcommand{\RReducesTo}{{\longrightarrow\!\!\!\!\!\rightarrow}}

\newcommand{\eqbe}{=_{\beta\eta}}
\def\lkbec/{$\lambda\mathbf{K}\beta\eta$-calculus}
\def\Schon/{Sch\"onfinkel}
\renewcommand{\And}{\wedge}

\newcommand{\Intersect}{\cap}

\newcommand{\mfbox}[1]{\mbox{\fbox{$#1$}}}
\newcommand{\aset}[1]{\mathsf{#1}}
\newcommand{\FreeVars}[1]{\aset{FreeVars}\!\left(#1\right)}
\newcommand{\Vars}{\aset{Vars}}
\newcommand{\vsc}[3]{{#1}_{#2}, \ldots, {#1}_{#3}}
\newcommand{\vs}[3]{{#1}_{#2} \cdots {#1}_{#3}}
\newcommand{\bbrack}[1]{\left\llbracket{#1}\right\rrbracket}
\newcommand{\ikbcs}{\comb I, \comb K, \comb B, \comb C, \comb S}
\newcommand{\vikbcs}{\vI, \vK, \vB, \vC, \vS}
\newcommand{\ag}{\mbox{\scriptsize{\textsc{ag}}}}
\newcommand{\variadic}[1]{{#1}_{\!\ag}}
\def\AGENallcaps/{ARITY-GENERIC}
\def\AGEN/{Arity-Generic}
\def\Agen/{Arity-generic}
\def\agen/{arity-generic}

\newcommand{\vPhi}{\variadic{\Phi}}
\newcommand{\vPsi}{\variadic{\Psi}}
\newcommand{\vI}{\vcomb I}
\newcommand{\vK}{\vcomb K}
\newcommand{\vB}{\vcomb B}
\newcommand{\vC}{\vcomb C}
\newcommand{\vS}{\vcomb S}
\newcommand{\vsig}{\variadic{\sigma}}
\newcommand{\vpi}{\variadic{\pi}}
\newcommand{\vNtupMaker}{\variadic{\NtupMaker}}
\newcommand{\vMap}{\variadic{\mathrm{Map}}}
\def\fpc/{fixed-point combinator}
\def\ms/{Sch\"onfinkel}
\def\lt/{$\lambda$-term}
\def\le/{$\lambda$-expression}
\def\cn/{Church numeral}
\def\lc/{$\lambda$-calculus}
\def\cl/{combinatory logic}
\newcommand{\eeta}{{\setbox0=\hbox{$=_{\eta}$}\setbox1=\hbox{$=$}\wd0=\wd1\mbox{\box0}}}
\newcommand{\comb}[1]{\mathbf{#1}}
\newcommand{\vcomb}[1]{\variadic{\mathbf{#1}}}
\newcommand{\Pred}{P^-}
\newcommand{\Succ}{S^+}
\newcommand{\Monus} 
  {{\mbox{\,\hbox{$-$}\hspace{-0.5em}\raisebox{0.2em}{\hbox{$\cdot$}}}\;}}
\newcommand{\Zero}{\comb{Zero?}}
\newcommand{\Ntup}[1]{\left\langle{#1}\right\rangle}
\newcommand{\NtupMaker}{\Ntup{\mbox{\textvisiblespace}, \ldots,\mbox{\textvisiblespace}}}
\newcommand{\Yname}[1]{\mathrm{Y}_{\!\mbox{\tiny\textsc{#1}}}}
\newcommand{\Ycurry}{\Yname{Curry}}
\newcommand{\Yturing}{\Yname{Turing}}
\newcommand{\Set}[1]{\left\{{#1}\right\}}

\newcommand{\alias}[2]{\underbrace{#1}_{\!\!\!\!\!\!\mbox{\small aliased by $#2$}\!\!\!\!\!\!}}
\newcommand{\unalias}[2]{\underbrace{#1}_{\!\!\!\!\!\!\mbox{\small un-aliasing $#2$}\!\!\!\!\!\!}}

\def\ver#1#2@{
  \def\loop##1{\ifx##1\emptybox\let\loop=\relax\else
    \\{##1}\fi\loop
  }
  \begingroup\!\!\!\begin{array}[t]{l@{}}#1\loop#2\emptybox\end{array}\endgroup
}

\allowdisplaybreaks

\title[Ellipses and Lambda Definability]{Ellipses and Lambda Definability}
\author[M.~Goldberg]{Mayer Goldberg}
\address{Department of Computer Science, Ben-Gurion University, Beer Sheva 8410501, Israel.}
\email{gmayer@cs.bgu.ac.il}

\begin{document}

\begin{abstract}
  \noindent Ellipses are a meta-linguistic notation for denoting terms
  the size of which are specified by a meta-variable that ranges over
  the natural numbers. In this work, we present a systematic approach
  for encoding such meta-expressions in the \lc/, without ellipses:
  Terms that are parameterized by meta-variables are replaced with
  corresponding $\lambda$-abstractions over actual variables. We call
  such \lt/s \textit{\agen/}. Concrete terms, for particular choices
  of the parameterizing variable are obtained by applying an \agen/
  \lt/ to the corresponding numeral, obviating the need to use
  ellipses.
  
  For example, to find the multiple fixed points of $n$ equations, $n$
  different \lt/s are needed, every one of which is indexed by two
  meta-variables, and defined using three levels of ellipses. A single
  \agen/ \labs/ that takes two Church numerals, one for the number of
  fixed-point equations, and one for their arity, replaces all these
  multiple fixed-point combinators. We show how to define \agengen/s
  of two historical \fpc/s, the first by Curry, and the second by
  Turing, for defining multiple fixed points. These historical \fpc/s
  are related by a construction due to \boehm/: We show that likewise,
  their \agengen/s are related by an \agengen/ of \boehm/'s
  construction.

  We further demonstrate this approach to \agen/ \ldef/ with
  additional \lt/s that create, project, extend, reverse, and map over
  ordered $n$-tuples, as well as an \agen/ generator for one-point
  bases.

 \end{abstract}


\maketitle


\section{Introduction}

\subsection{Motivation}
\label{ssec:motivation}

This work is concerned with \lt/s that are written using the \linebreak
meta-language of ellipses: Terms such as, for example, the ordered
$n$-tuple \linebreak maker: $\lambda \vs x1n \sigma.(\sigma\ \vs x1n$). As the
use of ellipses indicates, the syntax for such \lt/s is described
\textit{for any given $n$}, in the meta-language of the \lc/, i.e., in
the language in which we describe the syntax of \lt/s. The index $n$
is thus a meta-variable. It is only after we have picked a natural
number for $n$, that we can write down an actual \lt/, and it will be
``hard-coded'' for that specific $n$. For example, the ordered
$5$-tuple maker is defined as $\lambda x_1 x_2 x_3 x_4 x_5 \sigma
. (\sigma\ x_1\ x_2\ x_3\ x_4\ x_5)$, can be written without ellipses,
and is ``hard-coded'' for $n = 5$.  But what if we want $n$, which
determines the syntactic structure of the \lt/, to be an argument in
the language of the \lc/: How do we go from a \lt/ whose syntax is
indexed or parameterized by a meta-variable over the natural numbers
in the meta-language of the \lc/ to a corresponding \lt/ parameterized
by a Church numeral?

In this work, we present a systematic approach for encoding terms
whose syntax is parameterized by a meta-variable and written using
ellipses, to \lt/s that take a Church numeral $c_n$ as an argument,
and return the corresponding \lt/ for that given $n$. We call such
\lt/s \textit{\agen/}, following the work of Weirich and Casinghino on
\textit{\AGEN/ Datatype-Generic Programming}~\cite{Weirich2010}. When
we speak of an \agen/ \lt/ $\variadic{E}$, we require two things:
\begin{enumerate}
\item We have in mind an $n$-ary term $E_n$ in the meta-language of
  the \lc/, that is parameterized by a meta-variable $n \in
  \mathbb{N}$. For any specific value of $n$, $E_n$ is a \lt/: $E_1,
  E_3$, etc., are all \lt/s.
\item For all $n \in \mathbb{N}$, $(\variadic{E}\ c_n) \eqbe E_n$.
\end{enumerate}

\subsection{Overview}

In Combinatory Logic, bases provide a standard approach to
constructing inductively larger combinators from smaller
combinators. We follow this approach by augmenting the standard of
$\Set{\ikbcs}$ basis introduced by \Schon/~\cite{Schoenfinkel1924},
Curry~\cite{Curry1958,Curry1972}, Turner~\cite{Turner1979a}, and many
others, with \agengen/s $\vK, \vS$ of the respective $\comb K, \comb
S$ combinators. We then encode $\vK, \vS$ in terms of
$\Set{\ikbcs}$~(Section~\ref{sec:AGext}). $\vK, \vS$ can then be used
to encode straightforwardly those parts of the term that use ellipses
using an \agengen/ of the \ba/ algorithm for the $\Set{\comb K, \comb
  S}$ basis.

{\sloppy
In principle, we could have stopped at this point, since $\Set{\ikbcs,
  \vK, \vS}$ would already be sufficient to encode any \agen/ term. We chose,
however, to use $\vK, \vS$ to define $\vB, \vC$, which are the \agen/
generalizations of $\comb B, \comb C$, because Turner's \textit{\ba/
  algorithm} for the basis $\Set{\ikbcs}$ extends naturally to the basis
$\Set{\ikbcs, \vikbcs}$. This extended
algorithm~(Section~\ref{sec:extending}) maintains the simplicity of
Turner's original algorithm, and generates compact encodings for
\agen/\ \lt/s.
}

The second part of this work~(Section~\ref{sec:induct}) demonstrates
how the new basis can be used to encode interesting \agen/ \lt/s, such
as multiple fixed-point combinators.

\subsection{Terminology, notation and list of combinators}

For background material on the \lc/, we refer the reader to Church's
original book on the \lc/, \textit{The Calculi of Lambda
  Conversion}~\cite{Church1941}, Curry's two volumes
\textit{Combinatory Logic I, II}~\cite{Curry1958,Curry1972}, and
Barendregt's encyclopedic textbook, \textit{The Lambda Calculus: Its
  Syntax and Semantics}~\cite{Barendregt1984}. Here we briefly list
the \lt/s and notation used throughout this work.

\noindent\begin{longtable}{p{0.18\textwidth}p{0.35\textwidth}p{0.40\textwidth}}
$\comb I$ & 
$\lambda x . x$
&
\textit{Identit{\"a}tsfunktion}~\cite{Schoenfinkel1924}
\\
$\comb K$ &
$\lambda x y.x$
&
\textit{Konstanzfunktion}~\cite{Schoenfinkel1924}
\\
$\comb B$ &
$\lambda xyz.(x\ (y\ z))$ &
\textit{Zusammensetzungsfunktion}~\cite{Schoenfinkel1924}
\\
$\comb C$ &
$\lambda xyz.(x\ z\ y)$ &
\textit{Vertauschungsfunktion}~\cite{Schoenfinkel1924}
\\
$\comb S$ &
$\lambda xyz.(x\ z\ (y\ z))$ &
\textit{Verschmelzungsfunktion}~\cite{Schoenfinkel1924}
\\[0.5em]
$\Ntup{x_1, \ldots, x_n}$ &
$\lambda \sigma.(\sigma\ \vs x1n)$ &
Ordered $n$-tuple~\cite{Barendregt1984}\\
$\NtupMaker_n$ &
$\lambda \vs x1n \sigma.(\sigma\ \vs x1n)$ &
Ordered $n$-tuple maker~\cite{Goldberg-lambda-calculus-tutorial}\\
$\sigma_k^n$ &
$\lambda x_1\cdots x_n.x_k$
&
Selector: Returns the $k$-th of $n$ arguments~\cite{Barendregt1984}
\\
$\pi_k^n$ &
$\lambda x.(x\ \sigma_k^n)$
&
Projection: Returns the $k$-th projection of an ordered $n$-tuple~\cite{Barendregt1984}
\\[0.5em]
$c_n$ &
$\lambda sz.(\underbrace{s\ (\cdots(s}_
  {\!\!\!\!\mbox{$n$ times}\!\!\!\!}z)\cdots))$
&
The $n$-th \cn/ \cite{Church1941}
\\
$\Succ$ &
$\lambda nsz.(s\ (n\ s\ z))$
&
Computes the \textit{successor} on Church numerals~\cite{Church1941}
\\
$+$ &
$\lambda a b.(b\ \Succ\ a)$ &
Computes addition on \cn/s~\cite{Church1941}
\\
$\Pred$ &
$\lambda n.(\pi_1^2\ (n\ 
\ver{(\lambda p.\langle\ver{(\pi_2^2\ p),}
                           {(\Succ\ (\pi_2^2\ p))\rangle)}@}
    {\Ntup{c_0, c_0}))}@$
&
Computes the \textit{predecessor} on \cn/s~\cite{Goldberg-lambda-calculus-tutorial}, following Kleene's construction for the $\lambda\comb I\beta\eta$-calculus~\cite{Kleene1935a}
\\
$\Monus$ &
$\lambda ab.(b\ \Pred\ a)$
&
Computes the \textit{monus} function on \cn/s~\cite{Church1941}
\\[0.5em]
$\comb{False}$ &
$\lambda xy.y$ &
The Boolean value \textit{False}~\cite{Barendregt1984}
\\
$\comb{True}$ &
$\lambda xy.x$ &
The Boolean value \textit{True}~\cite{Barendregt1984}
\\
$\Zero$ &
$\lambda n.(n\ (\lambda x.\comb{False})\ \comb{True})$
&
Computes the \textit{zero-predicate} on \cn/s
\\
\end{longtable}

For any \lt/ $P$, the set of variables that occur freely in $P$ is
denoted by $\FreeVars P$. The $\equiv$ symbol denotes identity modulo
$\alpha$-conversion, the symbol $\RReducesTo$ denotes reflexive and
transitive closure of the $\beta\eta$ relation, The $=_{\beta}$ symbol
denotes the equivalence relation induced by $\beta$-reduction. The
$=_{\eta}$ symbol denotes the equivalence relation induced by
$\eta$-reduction. The symbol $\eqbe$, which is also abbreviated as
$=$, denotes the equivalence relation induced by the $\beta\eta$
relation.

The \textit{size} of a \lt/ $P$, denoted by $|P|$, is the length of
its abstract-syntax tree. For variable $\nu$, and \lt/s $P, Q$, we
have:
\begin{eqnarray*}
  | \nu | & = & 1 \\
  | \lambda \nu . P | & = & 1 + | P | \\
  | (P Q) | & = & 1 + | P | + | Q | \\
\end{eqnarray*}
The relationship between \lt/s $A, B$, and a function $f$, which maps the \lt/ $A$ to $B$ is denoted by $\Maps A f B$.

\subsection{The meta-language of ellipses}

The ellipsis is used extensively in the literature on the \lc/ and
\cl/: It appears in Church's original text on the
\lc/~\cite{Church1941}, in Curry's texts on
\cl/~\cite{Curry1958,Curry1972}, in Barendregt's text on the
\lc/~\cite{Barendregt1984}, and in many other books and articles.

As a meta-linguistic notational device, the ellipsis is very
economical, but the economy often hides subtlety and complexity. For
example, in the expression \[P \equiv \lambda x.\underbrace{(\Succ
  \cdots (\Succ}_{\mbox{$100$ times}}\ x)\cdots)\,,\] the ellipses serve
to abbreviate an expression that would otherwise be cumbersome to
write. Now consider the superficially-similar expression \[Q_n \equiv
\lambda x.\underbrace{(\Succ \cdots (\Succ}_{\mbox{$n$
    times}}\ x)\cdots )\,.\] For specific values of $n$, the expression
$Q_n$ is a \le/: $Q_1, Q_{23}, Q_{100}$, etc., are all \le/s, and in
fact, $Q_{100} \equiv P$. However, $Q_n$ is not a \le/:
Linguistically, $n$ is a meta-variable in the meta-language of the
\lc/, and so $Q_n$ is rather a meta-expression.

Would it be possible to define a \le/ that would, in some sense,
``capture the essence'' of $Q_n$? Since we use Church numerals in this
paper, and since Church numerals are abstractions over the iterated
composition of a function, it seems reasonable to argue that the
expression $R = \lambda n.\lambda x.(n\ \Succ\ x) =_{\eta} \lambda
n.(n\ \Succ)$ is our candidate: It takes a Church numeral $n$ as an
argument, and returns a function that applies to its argument the
$n$-th composition of $\Succ$. The relationship between $Q_n$ and $R$
is given by $Q_n = (R\ c_n)$. We can use this relationship, to replace
a meta-expression with a \le/ and a Church numeral, and in that sense,
``eliminate'' the use of ellipses.

In more complicated scenarios, ellipses and meta-variables can be
combined to hide even greater complexity. For example, in
Section~\ref{ssec:motivation}, we described the $n$-tuple maker:
$\NtupMaker_n = \lambda \vs x1n \sigma.(\sigma\ \vs x1n)$. Ellipses
now control the number of nested $\lambda$-abstrac\-tions, and the
number of left-associated applications. How can these ellipses be
eliminated? The ``interface'' to such a term, which we call $\variadic
\NtupMaker$ would take a Church numeral $c_n$, and satisfy the
relationship $(\variadic \NtupMaker\ c_n) = \NtupMaker_n$.

Sections~\hbox{\ref{sec:AGext},~\ref{sec:extending}, and
  \ref{sec:induct}} explore how all meta-linguistic ellipses can be
removed from expressions in the meta-language of the \lc/. Put
otherwise, the \lc/ is sufficiently expressive so as to make the use
of meta-linguistic ellipses unnecessary, even if they are still used
as a matter of convenience.

\section{\Agen/ generalizations of the $\Set{\comb I, \comb K, \comb B, \comb C, \comb S}$ basis}
\label{sec:AGext}

Our goal is to define \agen/ versions of $\comb I, \comb K, \comb B,
\comb C, \comb S$ combinators, which form the \agen/ part of a basis
for \agen/ \le/s.

\subsection{The \agen/ $\comb K$ combinator}

The $\comb K$ combinator, defined as $\lambda p x.p$, abstracts
a variable $x$ over an expression in which $x$ does not occur
free. The $n$-ary generalization of $\comb K$ abstracts $n$
variables, and is given by:
\begin{eqnarray*}
  \comb{K}_n & \equiv & \lambda p \vs x1n . p
\end{eqnarray*}
Notice that $\comb K$ abstracts a single unused variable over its
argument. Hence we may write:
\begin{eqnarray*}
\comb K_n & = & (c_n\ \comb K)
\end{eqnarray*}

We now define $\vK$ as follows:
\begin{eqnarray*}
  \vK
  & \equiv & \lambda n . (n\ \comb K)
\end{eqnarray*}
This definition satisfies the requirement that $(\vK\ c_n) =
\comb{K}_n$. Also note that $\comb K_0 = \comb I$, and $\comb{K}_1 =
\comb K$.

\subsection{The \agen/ $\comb S$ combinator}

The $\comb S$ combinator, defined as $\lambda p q x.(p\ x\ (q\ x))$,
abstracts a variable $x$ over an application of two expressions, where
$x$ occurs free in both expressions. The $n$-ary generalization of
$\comb S$ abstracts $n$ variables, and is given
by:\footnote{Curry~\cite[page~169]{Curry1958} uses the symbol $\comb
  S_n$ to denote the following generalization of $\comb S$, which is
  different from our own:
  \begin{eqnarray*}
    \comb S_n^{\mbox{\tiny{Curry}}} & \equiv & 
      \lambda f g_1 \cdots g_n x . (f\ x\ (g_1\ x) \cdots (g_n\ x))
  \end{eqnarray*}
  Nevertheless, we think that our generalization fits better here,
  because of the way the relevant rule in Turner's \ba/ algorithm for
  the basis $\Set{\ikbcs}$ generalizes to our definition of $\comb
  S_n$.}
\begin{eqnarray*}
\comb{S}_n & \equiv &
\lambda p q x_1 \cdots x_n . (p\ x_1 \cdots x_n\ (q\ x_1 \cdots x_n))
\end{eqnarray*}
We describe $\comb S_{n+1}$ in terms of $\comb S_n$:
\begin{eqnarray*}
\comb S_{n+1}
& = & \lambda p q x_1 \cdots x_{n+1} . 
        (p\ x_1 \cdots x_{n+1}\ (q\ x_1 \cdots x_{n+1})) \\
& = & \lambda p q x_1 \cdots x_{n+1} .
        (\comb S\ (p\ x_1 \cdots x_n)\ (q\ x_1 \cdots x_n)\ x_{n+1}) \\
& \eeta & \lambda p q x_1 \cdots x_n . 
        (\comb S\ (p\ x_1 \cdots x_n)\ (q\ x_1 \cdots x_n)) \\
& = & \lambda p q x_1 \cdots x_n . 
        (\comb K_n\ \comb S\ x_1 \cdots x_n\ 
        (p\ x_1 \cdots x_n)\ (q\ x_1 \cdots x_n)) \\
& = & \lambda p q x_1 \cdots x_n . (\comb S_n\ (\comb K_n\ 
      \comb S)\ p\ x_1 \cdots x_n\ (q\ x_1 \cdots x_n)) \\
& = & \lambda p q x_1 \cdots x_n . (\comb S_n\ (\comb S_n\ 
      (\comb K_n\ \comb S)\ p)\ q\ x_1 \cdots x_n) \\ 
& \eeta & \lambda p q . (\comb S_n\ (\comb S_n\ 
      (\comb K_n\ \comb S)\ p)\ q) \\
& \eeta & \lambda p . (\comb S_n\ (\comb S_n\ 
      (\comb K_n\ \comb S)\ p)) \\
& = & \lambda p . (\comb B\ \comb S_n\ (\comb S_n\ 
      (\comb K_n\ \comb S))\ p) \\
& \eeta & (\comb B\ \comb S_n\ (\comb S_n\ (\comb K_n\ \comb S))) \\
& = & ((\lambda s.(\comb B\ s\ (s\ (\comb K_n\ \comb S))))\ \comb S_n) 
\end{eqnarray*}
The \lt/ $f$ that takes a Church numeral $c_n$, and maps $\Maps{\comb S_n} f {\comb S_{n+1}}$ is given by
\begin{eqnarray*}
  f & = & \lambda ns.((\lambda s.(\comb B\ s\ (s\ (\vK\ n\ \comb S))))\ s) \\
  & = & \lambda ns.(\comb B\ s\ (s\ (\vK\ n\ \comb S)))
\end{eqnarray*}
The \lt/ $g$ such that $\Maps {\Ntup{c_n, \comb S_n}} g {\Ntup{c_{n+1}, \comb S_{n+1}}}$ is given by:
\begin{eqnarray*}
g & = & 
\lambda p.\Ntup{(\Succ\ (\pi_1^2\ p)), (f\ (\pi_1^2\ p)\ (\pi_2^2\ p))}
\end{eqnarray*}
Note that $\comb S_0$ abstracts over $0$ arguments, so we have $\comb S_0 = \lambda pq.(pq) =_{\eta} \comb I$. We define $\vS$ by taking the $n$-th composition of $g$, applying it to $\Ntup{c_0, \comb S_0}$, and taking the second projection:
\begin{eqnarray*}
  \vS & \equiv &
  \lambda n.(\pi_2^2\ (n\ g\ \Ntup{c_0, \comb I}))
\end{eqnarray*}
This definition satisfies the requirement that $(\vS\ c_n) =
\comb{S}_n$. Also note that $\comb{S}_0 = \comb I$, and $\comb S_1 =
\comb S$.

\subsection{The \agen/ $\comb I$ combinator}

The $\comb I$-combinator is defined as $\lambda x.x$. The $n$-ary
generalization of $\comb I$ is
\begin{eqnarray*}
\comb I_n & \equiv & \lambda x_1 \cdots x_n.(x_1 \cdots x_n) \\
& \eeta & \comb I
\end{eqnarray*}
Since $\comb I_n =_{\eta} \comb I$, this case is trivial. It is
nevertheless necessary for completeness, to give the \agen/
extension of $\comb I_n$:
\begin{eqnarray*}
\variadic {\comb I} & \equiv & \lambda n . \comb I \\
& = & (\comb K\ \comb I)
\end{eqnarray*}

\noindent This definition trivially satisfies the requirement that
$(\variadic{\comb I}\ c_n) = \comb I_n$, as $\comb I_n = \comb I$
holds trivially for all $n \in \mathbb{N}$.

\subsection{The \agen/ $\comb B$ combinator}

The $\comb B$ combinator, defined as $\lambda pqx.(p\ (q\ x))$ abstracts a variable $x$ over an application of two expressions, where $x$ occurs free in the second expression. The $n$-ary generalization of $\comb B$ abstracts $n$ variables, and is given by:
\begin{eqnarray*}
\comb{B}_n 
& \equiv & \lambda p q x_1 \cdots x_n . (p\ (q\ x_1 \cdots x_n)) \\
& = & 
\lambda p q x_1 \cdots x_n .
(\comb K_n\ p\ x_1 \cdots x_n\ (q\ x_1 \cdots x_n)) \\
& = &
\lambda p q x_1 \cdots x_n .
(\comb S_n\ (\comb K_n\ p)\ q\ x_1 \cdots x_n) \\
& \eeta &
\lambda p . (\comb S_n\ (\comb K_n\ p)) \\
& = & 
\lambda p . (\comb B\ \comb S_n\ \comb K_n\ p) \\
& \eeta &
(\comb B\ \comb S_n\ \comb K_n)
\end{eqnarray*}
The \agen/ version of $\comb B$, written as $\vB$ takes $c_n$ and returns $\comb B_n$. We can define $\vB$ as follows:
\begin{eqnarray*}
\vB & \equiv &
\lambda n.(\comb B\ (\vS\ n)\ (\vK\ n))
\end{eqnarray*}
This definition satisfies the requirement that $(\vB\ c_n) = \comb
B_n$. Also note that $\comb B_0 = \comb I$, and $\comb B_1 = \comb B$.

\subsection{The \agen/ $\comb C$ combinator}

The $\comb C$ combinator, defined as $\lambda pqx.(p\ x\ q)$ abstracts a
variable $x$ over an application of two expressions, where $x$ occurs
free in the first expression. The $n$-ary generalization of $\comb C$
abstracts $n$ variables, and is given by:
\begin{eqnarray*}
\comb C_n 
& = &
\lambda p q x_1 \cdots x_n . (p\ x_1 \cdots x_n\ q) \\
& = &
\lambda p q x_1 \cdots x_n . 
(p\ x_1 \cdots x_n\ (\comb K_n\ q\ x_1 \cdots x_n)) \\ 
& = &
\lambda p q x_1 \cdots x_n .
(\comb S_n\ p\ (\comb K_n\ q)\ x_1 \cdots x_n) \\
& \eeta &
\lambda p q . 
(\comb S_n\ p\ (\comb K_n\ q)) \\
& = &
\lambda p q . 
(\comb B\ (\comb S_n\ p)\ \comb K_n\ q) \\
& = &
\lambda p q . 
(\comb B\ \comb B\ \comb S_n\ p\ \comb K_n\ q) \\
& = &
\lambda p q . 
(\comb C\ (\comb B\ \comb B\ \comb S_n)\ \comb K_n\ p\ q) \\
& \eeta &
(\comb C\ (\comb B\ \comb B\ \comb S_n)\ \comb K_n)
\end{eqnarray*}
The \agen/ version of $\comb C$, written as $\vC$ takes $c_n$ and returns $\comb C_n$. We can define the $\vC$ as follows:
\begin{eqnarray*}
\vC & \equiv &
\lambda n . (\comb C\ (\comb B\ \comb B\
(\vS\ n))\ (\vK\ n))
\end{eqnarray*}
This definition satisfies the requirement that $(\vC\ c_n) = \comb C_n$. Also note that $\comb C_0 = \comb I$, and $\comb C_1 = \comb C$.

\subsection{Summary and Conclusion}

We have introduced $n$-ary and \agengen/s of the combinators $\comb I, \comb K, \comb B, \comb C, \comb S$. These terms satisfy the property that for any $X \in \Set{\comb I, \comb K, \comb B, \comb C, \comb S}$, we have $(\variadic X\ c_n) = X_n$, and in particular $(\variadic X\ c_1) = X_1 = X$. 

Encoding an $n$-ary extension of a \lt/ parallels the case where $n = 1$, both in the steps as well as in the final encoding. For example, consider the parallel encoding of $\comb B$ and $\comb B_n$:
\[
\eqalign{
\comb B x y z 
&\ =\ x (y z) \cr
&\ =\ \comb K x z (y z) \cr
&\ =\ \comb S (\comb K x) y z  \cr
&\ =\ \comb K \comb S x (\comb K x) y z  \cr
&\ =\ \uline{\comb S (\comb K \comb S) \comb K} x y z 
}
\qquad
\eqalign{
\comb B_n x y \vs z1n
&\ =\  x (y \vs z1n) \cr
&\ =\  \comb K_n x \vs z1n (y \vs z1n) \cr
&\ =\  \comb S_n (\comb K_n x) y \vs z1n  \cr
&\ =\  \comb K \comb S_n x (\comb K_n x) y \vs z1n  \cr
&\ =\  \uuline{\comb S (\comb K \comb S_n) \comb K_n} x y \vs z1n 
}
\]\smallskip

\noindent Hence we obtain an alternative encoding for an \agen/ extension of $\comb B$ as follows:
\begin{eqnarray*}
\vB^{\mbox{\tiny{alt}}} & \equiv &
\lambda n.(\comb S\ (\comb K\ (\vS\ n))\ (\vK\ n))
\end{eqnarray*}
Similarly, consider the parallel encoding of $\comb C$ and $\comb C_n$:\smallskip
\[
\eqalign{
\comb C x y z 
&\ =\ x z y \cr
&\ =\ x z (\comb K y z) \cr
&\ =\ \comb S x (\comb K y) z \cr
&\ =\ \comb K (\comb S x) y (\comb K y) z \cr
&\ =\ \comb S (\comb K (\comb S x)) \comb K y z \cr
&\ =\ \comb S (\comb K \comb K x (\comb S x)) \comb K y z \cr
&\ =\ \comb K \comb S x (\comb S (\comb K \comb K) \comb S x) \comb K y z \cr
&\ =\ \comb S (\comb K \comb S) (\comb S (\comb K \comb K) \comb S) x \comb K y z \cr
&\ =\ \comb S (\comb K \comb S) (\comb S (\comb K \comb K) \comb S) x (\comb K \comb K x) y z \cr
&\ =\ \uline{\comb S (\comb S (\comb K \comb S) (\comb S (\comb K \comb K) \comb S)) (\comb K \comb K)} x y z
}\qquad
\eqalign{\llap{$\comb C_n x y \vs z1n$}
&\ =\ x \vs z1n y \cr
&\ =\ x \vs z1n (\comb K_n y \vs z1n) \cr
&\ =\ \comb S_n x (\comb K_n y) \vs z1n \cr
&\ =\ \comb K (\comb S_n x) y (\comb K_n y) \vs z1n \cr
&\ =\ \comb S (\comb K (\comb S_n x)) \comb K_n y \vs z1n \cr
&\ =\ \comb S (\comb K \comb K x (\comb S_n x)) \comb K_n y \vs z1n \cr
&\ =\ \comb K \comb S x (\comb S (\comb K \comb K) \comb S_n x) \comb K_n y \vs z1n \cr
&\ =\ \comb S (\comb K \comb S) (\comb S (\comb K \comb K) \comb S_n) x \comb K_n y \vs z1n \cr
&\ =\ \comb S (\comb K \comb S) (\comb S (\comb K \comb K) \comb S_n) x (\comb K \comb K_n x) y \vs z1n \cr
&\ =\  \uuline{\comb S (\comb S (\comb K \comb S) (\comb S (\comb K \comb K) \comb S_n)) (\comb K \comb K_n)} x y \vs z1n
}
\]\smallskip

\noindent Hence we obtain an alternative encoding for an \agen/ extension of $\comb C$ as follows:
\begin{eqnarray*}
\vC^{\mbox{\tiny{alt}}} & \equiv & 
\lambda n.(\comb S\ (\comb S\ (\comb K\ \comb S)\ (\comb S\ (\comb K\ \comb K) (\vS\ n)))\ (\comb K\ (\vK\ n)))
\end{eqnarray*}

\Agen/ \lt/s can be encoded directly using $\Set{\ikbcs, \vK, \vS}$
and Church numerals, similarly to how combinators are encoded using
$\Set{\comb K, \comb S}$, and we have done just that in encoding the
$\vB, \vC$ combinators. 

Our aim, however, was to extend the original $\Set{\ikbcs}$ basis
introduced by \Schon/, resulting in a more compact encoding, and in a
smaller number of derivation steps.

\section{Turner's \BA/ Algorithm}

A \ba/ algorithm is an algorithm for translating a \le/ into an
equivalent expression that is generated by some basis, an expression
that contains no $\lambda$-abstractions and no variables, and that is
written using applications of the terms of the given basis. Thus a
\ba/ algorithm is specific to a given basis. 

Turner's \ba/ algorithm~\cite{Turner1979a} is an algorithm for
translating \le/s into the $\Set{\ikbcs}$ basis. The algorithm,
denoted by \textit{double brackets} ($\bbrack{\cdot}$) is defined on
the structure of the argument, and is described in several cases:

\begin{longtable}{lll}
  \textbf{Original term} & \textbf{Condition} & \textbf{Rewrite} \\
  $M \in \Vars$ & & $M$ \\
  $M \in \Set{\ikbcs}$ & & $M$ \\
  $M = (P\ Q)$ & & $(\bbrack{P}\ \bbrack{Q})$ \\
  $M = \lambda x.\lambda y.P$ & & 
    $\bbrack{\lambda x.\bbrack{\lambda y.P}}$ \\
  $M = \lambda x.(P\ x)$ & $x \not\in \FreeVars{P}$ & $\bbrack{P}$ \\
  $M = \lambda x.P$ & $x \not\in \aset{FreeVars}(P)$ &
    $(\comb K\ \bbrack{P})$ \\
  $M = \lambda x.(P_x\ Q)$ & 
    $x \in \aset{FreeVars}(P_x),
    x \not\in \aset{FreeVars}(Q)$ &
  $(\comb C\ \bbrack{\lambda x.P_x}\ \bbrack{Q})$ \\
  $M = \lambda x.(P\ Q_x)$ & $x \not\in \aset{FreeVars}(P),
    x \in \aset{FreeVars}(Q_x)$ &
    $(\comb B\ \bbrack{P}\ \bbrack{\lambda x.Q_x})$ \\
  $M = \lambda x.(P_x\ Q_x)$ & $x \in \aset{FreeVars}(P_x),
    x \in \aset{FreeVars}(Q_x)$ &
    $(\comb S\ \bbrack{\lambda x.P_x}\ \bbrack{\lambda x.Q_x})$ \\
\end{longtable}

\noindent To give some intuition as to the r\^ole the different combinators of
the basis play in the algorithm, let us analyze just one single case:
Where $M = \lambda x.(P_x\ Q_x)$:
\begin{eqnarray*}
  \lambda x.(P_x\ Q_x) & = &
  (\underbrace{\mfbox{(\lambda pqx.(p\ x\ (q\ x)))}}_{\equiv \comb S}\ (\lambda x.P_x)\ (\lambda x.Q_x))
\end{eqnarray*}
Hence we have the rule that
\begin{eqnarray*}
  \bbrack{\lambda x.(P_x\ Q_x)} & = &
  (\comb S\ \bbrack{\lambda x.P_x}\ \bbrack{\lambda x.Q_x})
\end{eqnarray*}

\noindent The correctness of this algorithm is shown by induction on the
\textit{length} of the term, rather than by structural induction,
because, for example, while $|\lambda x.P_x | < | \lambda
x.(P_x\ Q_x)|$, clearly $\lambda x.P_x$ is not a sub-expression of
$\lambda x.(P_x\ Q_x)$, and the same holds for other cases in the
proof.

\noindent\textbf{Example: } We demonstrate the \ba/ algorithm by
applying it to $\Succ$: 
\begin{eqnarray*}
  \bbrack{S^+}
  & \equiv & \bbrack{\lambda abc.(b\ (a\ b\ c))} \\
  & = & \bbrack{\lambda a. \bbrack{\lambda b. \bbrack{\lambda c.(b\ (a\ b\ c))}}} \\
  & = & \bbrack{\lambda a. \bbrack{\lambda b. (\comb
      B\ \bbrack{b}\ \bbrack{\lambda c.(a\ b\ c)})}} \\
  & =_{\eta} & \bbrack{\lambda a. \bbrack{\lambda b. (\comb B\ b\ (a\ b))}} \\
  & = & \bbrack{\lambda a. (\comb S\ \bbrack{\lambda b.(\comb B\ b)}\ \bbrack{\lambda b.(a\ b)})} \\
  & =_{\eta} & \bbrack{\lambda a. (\comb S\ \comb B\ a)} \\
  & =_{\eta} & (\comb S\ \comb B)
\end{eqnarray*}

\section{Extending Turner's \BA/ Algorithm}
\label{sec:extending}

\subsection{Extending the rule for $\comb I$}

In Turner's original bracket-abstraction algorithm, the rule for $\comb I$ was a base case:
\begin{eqnarray*}
\bbrack{\lambda x . x} & = & \comb I
\end{eqnarray*}
The $n$-ary and \agengen/ of the rule for $\comb I$ abstracts $n$ variables $\vsc x1n$, and is also a base case:
\begin{eqnarray*}
\bbrack{\lambda \vs x1n.(\vs x1n)}
& = & \comb I_n \\
& = & (\vI\ c_n)
\end{eqnarray*}

\subsection{Extending the rule for $\comb K$}

Note that $(\comb K\ P) = ((\lambda p x.p)\ P) = \lambda x.P$. Accordingly, $\comb K$ is used in the original \ba/ algorithm to abstract a variable $x$ over an expression $P$, where $x \not\in \FreeVars P$:
\begin{eqnarray*}
\bbrack{\lambda x.P} & = & \left(
\comb K ~ \bbrack{P}
\right)
\end{eqnarray*}
The $n$-ary and \agengen/ of the rule for $\comb K$ allows for abstracting $n$ variables $\vsc x1n$ over an expression $P$, where $\Set{\vsc x1n} \Intersect \FreeVars P = \emptyset$:
\begin{eqnarray*}
\bbrack{\lambda x_1 \cdots x_n.P} & = & \left(
\comb K_n ~ \bbrack{P}
\right) \\
& = & \left(
\vK ~ c_n ~ \bbrack{P}
\right)
\end{eqnarray*}

\subsection{Extending the rule for $\comb B$}

Note that 
\begin{eqnarray*}
(\comb B\ P\ (\lambda x.Q_x)) 
& = & ((\lambda pqx.(p\ (q\ x)))\ P\ (\lambda x.Q_x)) \\
& = & \lambda x.(P\ Q_x)
\end{eqnarray*}
where $x \in \FreeVars{Q_x}$. Accordingly, $\comb B$ is used in the original \ba/ algorithm to abstract a variable $x$ over an application $(P\ Q_x)$, where $x \in \FreeVars{Q_x}$:
\begin{eqnarray*}
\bbrack{\lambda x.(P ~ Q_x)} & = & \left(
\comb B ~ \bbrack{P} ~ \bbrack{\lambda x.Q_x}
\right)
\end{eqnarray*}
The $n$-ary and \agengen/ of the rule for $\comb B$ allows for abstracting $n$ variables $\vsc x1n$ over an application $(P\ Q_{\vsc x1n})$, where $\Set{\vsc x1n} \Intersect \FreeVars P = \emptyset$ and $\Set{\vsc x1n} \subseteq \FreeVars{Q_{\vsc x1n}}$:
\begin{eqnarray*}
\bbrack{\lambda \vs x1n.(P ~ Q_{\vsc x1n})} & = & \left(
\comb B_n ~ \bbrack{P} ~ \bbrack{\lambda \vs x1n.Q_{\vsc x1n}}
\right) \\
& = & \left(
\vB ~ c_n ~ \bbrack{P} ~ \bbrack{\lambda \vs x1n.Q_{\vsc x1n}}
\right)
\end{eqnarray*}

\subsection{Extending the rule for $\comb C$}

Note that 
\begin{eqnarray*}
(\comb C\ (\lambda x.P_x)\ Q) & = & ((\lambda pqx.(p\ x\ q))\ (\lambda x.P_x)\ Q) \\
 & = & \lambda x.(P_x\ Q)
\end{eqnarray*}
where $x \in \FreeVars{P_x}$. Accordingly, $\comb C$ is used in the original \ba/ algorithm to abstract a variable $x$ over an application $(P_x\ Q)$, where $x \in \FreeVars{P_x}$:
\begin{eqnarray*}
\bbrack{\lambda x.(P_x\ Q)} & = & \left(
\comb C ~ \bbrack{\lambda x.P_x} ~ \bbrack{Q}
\right)
\end{eqnarray*}
The $n$-ary and \agengen/ of the rule for $\comb C$ allows for abstracting $n$ variables $\vsc x1n$ over an application $(P_{\vsc x1n}\ Q)$, where $\Set{\vsc x1n} \subseteq \FreeVars{P_{\vsc x1n}} \And \Set{\vsc x1n} \Intersect \FreeVars Q = \emptyset$:
\begin{eqnarray*}
\bbrack{\lambda \vs x1n.(P_{\vsc x1n} ~ Q)} & = & \left(
\comb C_n ~ \bbrack{\lambda \vs x1n . P_{\vsc x1n}} ~ \bbrack{Q}
\right) \\
& = & \left(
\vC ~ c_n ~ \bbrack{\lambda \vs x1n . P_{\vsc x1n}} ~ \bbrack{Q}
\right)
\end{eqnarray*}

\subsection{Extending the rule for $\comb S$}

Note that 
\begin{eqnarray*}
(\comb S\ (\lambda x.P_x)\ (\lambda x.Q_x)) 
& = & ((\lambda pqx.(p\ x\ (q\ x)))\ (\lambda x.P_x)\ (\lambda x.Q_x)) \\
& = & \lambda x.(P_x\ Q_x)
\end{eqnarray*}
where $x \in \FreeVars{P_x} \Intersect \FreeVars{Q_x}$. Accordingly, $\comb S$ is used in the original \ba/ algorithm to abstract a variable $x$ over an application $(P_x\ Q_x)$, where $x \in \FreeVars{P_x} \Intersect \FreeVars{Q_x}$:
\begin{eqnarray*}
\bbrack{\lambda x.(P_x ~ Q_x)} & = & \left(
\comb S ~ \bbrack{\lambda x.P_x} ~ \bbrack{\lambda x.Q_x}
\right)
\end{eqnarray*}
The $n$-ary and \agengen/ of the rule for $\comb S$ allows for abstracting $n$ variables $\vsc x1n$ over an application $(P_{\vsc x1n}\ Q_{\vsc x1n})$, where $\Set{\vsc x1n} \subseteq \FreeVars{P_{\vsc x1n}} \Intersect \FreeVars{Q_{\vsc x1n}}$:
\begin{eqnarray*}
\bbrack{\lambda \vs x1n.(P_{\vsc x1n} ~ Q_{\vsc x1n})} & = & (
\comb S_n ~ \ver{\bbrack{\lambda \vs x1n . P_{\vsc x1n}}}
                          {\bbrack{\lambda {\vs x1n} . Q_{\vsc x1n}})}@ \\
& = & (
\vS ~ c_n ~ \ver{\bbrack{\lambda \vs x1n . P_{\vsc x1n}}}
                        {\bbrack{\lambda {\vs x1n} . Q_{\vsc x1n}})}@
\end{eqnarray*}

\subsection{Summary and Conclusion}

In Turner's \ba/ algorithm, each of the combinators $\comb K, \comb B, \comb C, \comb S$ is used to encode an abstraction of a variable over an expression: The $\comb K$ combinator is used when the variable does not occur freely in the expression. The $\comb B, \comb C, \comb S$ combinators are used when the variable abstracts over an application of two expressions, and correspond to the situations where the given variable occurs freely in one or in both expressions. 

We extended Turner's \ba/ algorithm by introducing four additional
rules for $\vikbcs$, corresponding to the abstraction of a
\textit{sequence of variables} of an expression. The extended
algorithm shares the simplicity of Turner's original algorithm, and
generates compact encodings for \agen/\ \lt/s.

In those situations where $\Set{\vsc x1n} \Intersect \FreeVars{P} \neq \emptyset \And \Set{\vsc x1n} \not\subseteq \FreeVars{P}$, we can use the $\comb K$-introduction rule to obtain from $P$ a $\beta$-equal expression $P'$ for which $\Set{\vsc x1n} \subseteq \FreeVars{P'}$.

\begin{prop} For any $n$-ary \le/ $\mathcal{E}_n$
that is written with ellipses, a corresponding \agen/ \le/
$\variadic{\mathcal{E}}$ can be defined, such that for any natural
number $n$, we have $(\variadic{\mathcal{E}}\ c_n) = \mathcal{E}_n$.
\end{prop}

\indent\textit{Sketch of proof}: By induction on the length of
$\mathcal{E}_n$, a corresponding rule can be applied in the extended
algorithm, so that the rewritten expression is \agen/ and satisfies
the above relation to $\mathcal{E}_n$. \qed

\noindent\textbf{Example: } Church~\cite{Church1941} introduces the \le/ $\comb D = \lambda x.(x\ x)$, which is encoded via Turner's algorithm as $(\comb S\ \comb I\ \comb I)$. How would the $n$-ary and \agen/ extensions be encoded?

\indent\textit{The $n$-ary extension}: 
\begin{eqnarray*}
\bbrack{\comb D_n} 
& = & \bbrack{\lambda \vs x1n.(\vs x1n\ (\vs x1n))} \\
& = & (\comb S_n\ \bbrack{\lambda \vs x1n.(\vs x1n)}\ 
                  \bbrack{\lambda \vs x1n.(\vs x1n)}) \\
& = & (\comb S_n\ \comb I_n\ \comb I_n)
\end{eqnarray*}

\indent\textit{The \agen/ extension}: 
\begin{eqnarray*}
\variadic{\comb D} & = & \lambda n.(\vS\ n\ (\vI\ n)\ (\vI\ n))
\end{eqnarray*}
So as we can see, the extended basis $\Set{\ikbcs, \vikbcs}$ provides a natural extension of the original basis for encoding \agen/ \le/s.

\section{$n$-ary and \agen/ expressions}
\label{sec:induct}

\subsection{The \agen/ selector combinators}
\label{ssec:sel}

The \textit{selector} combinators return one of their arguments. For $n, k$ such that $0 \leq k \leq n$, the selector that returns the $k$-th of its $n+1$ arguments is defined as follows:
\begin{eqnarray*}
  \sigma_k^n 
  & \equiv & \lambda x_0 \cdots x_n . x_k
\end{eqnarray*}
An \agen/ version of the selector, which we write as $\vsig$, would take \cn/s $k, n$ and return $\sigma_k^n$. We generate $\sigma_k^n$ in two states: First, we generate a selector in which only the first argument is returned:
\begin{eqnarray*}
\lambda x_k \cdots x_n.x_k
\end{eqnarray*}
We then tag on $k$ additional abstractions. 

Suppose we have $P, Q$ that are defined as follows:
\begin{eqnarray*}
P & = & \lambda x_0 \cdots x_n . x_0 \\
Q & = & \lambda x_0 \cdots x_n x_{n+1} . x_0 
\end{eqnarray*}
We define the \lt/ $f$ to map $\Maps P f Q$ for all $n$. The relationship between $P$ and $Q$ is given by $Q = \lambda x.(P\ (\lambda z.x))$, and so:
\begin{eqnarray*}
f & = & \lambda px.(p\ (\lambda z.x))
\end{eqnarray*}
We use $f$ to generate $\lambda x_k \cdots x_n.x_k$ by applying the $(n-k)$-th composition of $f$ to the identity combinator $\comb I$. From this we obtain $\sigma_k^n$ by $k+1$ applications of $\comb K$. We can now define $\vsig$ as follows:
\begin{eqnarray*}
\vsig 
& \equiv & \lambda k n.(\Pred\ k\ \comb K\ (\Monus\ n\ k\ f\ \comb I))
\end{eqnarray*}
This definition satisfies the requirement that $(\vsig\ c_k\ c_n) =
\sigma_k^n$.

\subsection{The \agen/ projections}

The \textit{projection} combinators take an $n$-tuple and return the respective projection:
\begin{eqnarray*}
  (\pi_k^n\ \Ntup{\vsc x1n}) & = & x_k
\end{eqnarray*}
The standard way of defining projections is to take an $n$-tuple and apply it to the corresponding selector:
\begin{eqnarray*}
  \pi_k^n & \equiv & \lambda x . (x\ \sigma_k^n)
\end{eqnarray*}

The definition of the \agen/ extension $\vpi$ can be written in terms
of $\vsig$:
\begin{eqnarray*}
  \vpi 
  & \equiv & \lambda k n x . (x\ (\vsig\ k\ n)) 
\end{eqnarray*}
This definition satisfies the requirement that $(\vpi\ c_k\ c_n) =
\pi_k^n$.

\subsection{The \agen/, ordered $n$-tuple maker}
\label{ssec:nTupleMaker}

In his textbook \textit{The Lambda Calculus: Its Syntax and
  Semantics}~\cite[pages~133-134]{Barendregt1984}, Barendregt
introduces one of the standard constructions for
$n$-tuples\footnote{This construction appears, for ordered pairs and
  triples, in Church's book \textit{The Calculi of Lambda
    Conversion}~\cite{Church1941}.}:
\begin{eqnarray*}
\Ntup{E_1, \ldots, E_n} & \equiv &
\lambda z . (z\ E_1 \cdots E_n)
\end{eqnarray*}
The ordered $n$-tuple maker $\NtupMaker_n$ takes $n$ \lt/s and returns their ordered tuple. Although most texts on the \lc/ use it implicitly by using ordered tuples as generalizations to the syntax of the \lc/, it is easily definable:
\begin{eqnarray*}
\NtupMaker_n & \equiv & \lambda x_1 \cdots x_n . \Ntup{x_1, \ldots, x_n} \\
& = & \lambda x_1 \cdots x_n z . (z\ x_1\ \cdots x_n)
\end{eqnarray*}
We wish to define the \agengen/ of the $n$-tuple maker $\vNtupMaker$, such that: 
\begin{eqnarray*}
(\vNtupMaker\ c_n) & = & \NtupMaker_n
\end{eqnarray*}
We relate $\NtupMaker_n$ with $\NtupMaker_{n+1}$ as follows:
\begin{eqnarray*}
\NtupMaker_{n+1} 
& = & \lambda x_1 \cdots x_n x_{n+1} z.(z\ x_1 \cdots x_n\ x_{n+1}) \\
& = & \lambda x_1 \cdots x_n x_{n+1} z.
      (\NtupMaker_n\ x_1 \cdots x_n\ z\ x_{n+1}) \\
& = & \lambda x_1 \cdots x_n x_{n+1} z.
      (\comb C\ (\NtupMaker_n\ x_1 \cdots x_n)\ x_{n+1}\ z) \\
& \eeta & \lambda x_1 \cdots x_n.(\comb C\ (\NtupMaker_n\ x_1 \cdots x_n)) \\
& = & \lambda x_1 \cdots x_n.(\vB\ \comb C\ \NtupMaker_n\ x_1 \cdots x_n) \\
& \eeta & (\vB\ \comb C\ \NtupMaker_n)
\end{eqnarray*}
Using this relation, we define the \lt/ $f$ to map $\Maps {\Ntup{c_n, \NtupMaker_n}} f {\Ntup{c_{n+1}, \NtupMaker_{n+1}}}$ for all $n$:
\begin{eqnarray*}
f & = & 
\lambda p.\langle
\ver{(\Succ\ (\pi_1^2\ p)),}
    {(\vB\ (\pi_1^2\ p)\ \comb C\ (\pi_2^2\ p))\rangle}@
\end{eqnarray*}
Notice that $\NtupMaker_0 = \comb I$, so we can obtain $\NtupMaker_n$ by applying the $n$-th composition of $f$ to $\Ntup{c_0, \comb I}$ . We define $\vNtupMaker$ as follows:
\begin{eqnarray*}
\vNtupMaker & \equiv & 
\lambda n.(\pi_2^2\ (n\ f\ \Ntup{c_0, \comb I}))
\end{eqnarray*}

This definition satisfies the requirement that $(\vNtupMaker\ c_n) =
\NtupMaker_n$, so for example, $(\vNtupMaker\ c_3\ E_1\ E_2\ E_3) =
\Ntup{E_1, E_2, E_3}$.

The task of defining the $\NtupMaker_n$ combinator is given as an exercise in the author's course notes on the \lc/~\cite{Goldberg-lambda-calculus-tutorial}, where the combinator is referred to as \textit{malloc}, in a tongue-in-cheek reference to the C library function for allocating blocks of memory. 

\subsection{Applying \lt/s}

A useful property of our representation of ordered $n$-tuples, is that
it gives us \textit{left-associated} applications immediately:
\begin{eqnarray*}
  (\Ntup{E_1, \ldots, E_n}\ P) & = &
  (P\ E_1 \cdots E_n)
\end{eqnarray*}
This behavior can be used to apply some expression to its arguments, where these arguments are passed in an $n$-tuple, in much the same way as the \texttt{apply} procedure in LISP~\cite{McCarthy1962}, which takes a procedure and a list of arguments, and applies the procedure to these arguments. One notable difference though, is that the \lc/ does not have a notion of \agen/ procedures, and so the procedure we wish to apply must ``know'' now many arguments to expect. We can thus define:
\begin{eqnarray*}
\mathrm{Apply} & \equiv & 
\lambda f v.(v\ f)
\end{eqnarray*}

Because functions in the \lc/ are Curried, and therefore applications
associate to the left, $\mathrm{Apply}$ combinator proides for
\textit{left-associated} applications. For \textit{right-associated}
applications, we would like to have an \agen/ version of the following
$n$-ary \lt/:
\begin{eqnarray*}
\mathrm{RightApplicator}_n & \equiv &
\lambda x_1 \cdots x_n z . (x_1\ (x_2 \cdots (x_n\ z) \cdots))
\end{eqnarray*}
We begin by writing $\mathrm{RightApplicator}_{n+1}$ in terms of
$\mathrm{RightApplicator}_n$: 
\begin{eqnarray*}
\mathrm{RightApplicator}_{n+1} & = &
\lambda x_1 \cdots x_n x_{n+1} z . 
(\mathrm{RightApplicator}_n\ x_1 \cdots x_n\ (x_{n+1}\ z)) \\
& = &
\lambda x_1 \cdots x_n x_{n+1} z . 
(\comb B\ (\mathrm{RightApplicator}_n\ x_1 \cdots x_n)\ x_{n+1}\ z) \\
& = &
\lambda x_1 \cdots x_n x_{n+1} z . 
(\comb B_n\ \comb B\ \mathrm{RightApplicator}_n\ x_1 \cdots x_n\ x_{n+1}\ z) \\
& \eeta &
(\comb B_n\ \comb B\ \mathrm{RightApplicator}_n)
\end{eqnarray*}
The \lt/ $f$ such that $\Maps {\Ntup{c_n, \mathrm{RightApplicator}_n}} f {\Ntup{c_{n+1}, \mathrm{RightApplicator}_{n+1}}}$ is given by:
\begin{eqnarray*}
f & \equiv &
\lambda p.\langle
\ver{(\Succ\ (\pi_1^2\ p)),}
    {(\vB\ (\pi_1^2\ p)\ \comb B\ (\pi_2^2\ p))\rangle}@
\end{eqnarray*}
Notice that $\mathrm{RightApplicator}_0 = \lambda z.z = \comb I$, so
we can obtain $\mathrm{RightApplicator}_n$ by applying the $n$-th
composition of $f$ to $\Ntup{c_0, \comb I}$. We define $\variadic{\mathrm{RightApplicator}}$ as follows:
\begin{eqnarray*}
\variadic{\mathrm{RightApplicator}} & \equiv &
\lambda n.(\pi_2^2\ (n\ f\ \Ntup{c_0, \comb I}))
\end{eqnarray*}
This definition satisfies the requirement that
$(\variadic{\mathrm{RightApplicator}}\ c_n) =
\mathrm{RightApplicator}_n$.

\subsection{Extending $n$-tuples}

Applying $\NtupMaker_{n+1}$ to $n$ arguments results in a \lt/ that takes an argument and returns an $n+1$-tuple, in which the given argument is the $n+1$-st projection. We use this fact to extend an $n$-tuple by an additional $n+1$-st element:
\begin{eqnarray*}
\variadic{\mathrm{Extend}} & \equiv & 
\lambda n v a.(v\ (\vNtupMaker\ (\Succ\ n))\ a)
\end{eqnarray*}
We can use it as follows:
\begin{eqnarray*}
(\variadic{\mathrm{Extend}}\ c_n\ \Ntup{x_1, \ldots, x_n}\ x_{n+1}) & = &
\Ntup{x_1, \ldots, x_n, x_{n+1}}
\end{eqnarray*}

Similarly, we can define the \lt/ $\mathrm{Catenate}$, for creating an
$n+k$-tuple given an $n$-tuple and a $k$-tuple:
\begin{eqnarray*}
\mathrm{Catenate} & \equiv &
\lambda n v k w . (w\ (v\ (\vNtupMaker\ (+\ n\ k))))
\end{eqnarray*}
For example:
\begin{eqnarray*}
(\mathrm{Catenate}\ c_3\ \Ntup{E_1, E_2, E_3}\ c_2\ \Ntup{F_1, F_2}) 
& = & \Ntup{E_1, E_2, E_3, F_1, F_2}
\end{eqnarray*}

\subsection{Iota}

When working with indexed expressions, it is convenient to have the \textit{iota}-function (written as the Greek letter $\iota$, and pronounced ``yota''), which maps the number $N$ to the vector $\Ntup{0, \ldots, N-1}$. Iota was introduced by Kenneth Iverson first in the APL notation~\cite{Iverson1962}, and then in the APL programming language~\cite{Pakin1972}.

We implement the $\iota$ combinator to take a Church numeral $c_n$ and return the ordered $n$-tuple $\Ntup{c_0, \ldots, c_{n-1}}$. Given the standard definition of ordered $n$-tuples, it is natural to define $(\iota\ c_0) = \comb I$.

We know that
\begin{eqnarray*}
  (\iota\ c_n) & = & \Ntup{c_0, \ldots, c_{n-1}} \\
  & = & \lambda z . (z\ c_0 \cdots c_{n-1}\ c_n)
\end{eqnarray*}
So the \lt/ $f$ such that 
\begin{eqnarray*}
  (\iota\ c_n)
  & \stackrel{(f\ c_n)}{\Longrightarrow} & (\iota\ c_{n+1})
\end{eqnarray*}
can be characterized as follows:
\begin{eqnarray*}
(\lambda z.(z\ c_0 \cdots c_{n-1})) 
& \stackrel{(f\ c_n)}{\Longrightarrow} & 
(\lambda z . (z\ c_0 \cdots c_{n-1}\ c_n))
\end{eqnarray*}
We define $f$ as follows:
\begin{eqnarray*}
f & = & \lambda n i z . (i\ z\ n)
\end{eqnarray*}
The \lt/ $g$ such that
\begin{eqnarray*}
\Ntup{n, i} 
& \stackrel{g}{\Longrightarrow} & 
\Ntup{(\Succ\ n), (f\ n\ i)}
\end{eqnarray*}
is defined as follows:
\begin{eqnarray*}
g & = &
\lambda p . \langle
\ver{(\Succ\ (\pi_1^2\ p)),}
    {(f\ (\pi_1^2\ p)\ (\pi_2^2\ p))\rangle}@
\end{eqnarray*}
We now define $\iota$ as follows:
\begin{eqnarray*}
\iota & \equiv &
\lambda n . (\pi_2^2\ (n\ g\ \Ntup{c_0, \comb I}))
\end{eqnarray*}

This definition satisfies the requirement that $(\iota\ c_{n+1}) =
\lambda z.(z\ c_0 \cdots c_n)$.

\subsection{Reversing}

It is often useful to be able to reverse the arguments to a function
or an $n$-tuple. We can define an $n$-ary reversal combinator $R_n$ as
follows: 
\begin{eqnarray*}
R_n & \equiv &
\lambda x_1 \cdots x_n w . (w\ x_n \cdots x_1)
\end{eqnarray*}
$R_n$ can be used in two ways: 
\begin{enumerate}
\item We can use it to reverse an ordered $n$ tuple:
  \begin{eqnarray*}
    (\Ntup{E_1, \ldots, E_n}\ R_n) & = &
    \Ntup{E_n, \ldots, E_1}
  \end{eqnarray*}
\item We can use it to take $n$ arguments are return their $n$-tuple,
  in reverse order:
  \begin{eqnarray*}
    (R_n\ E_1 \cdots E_n) & = & \Ntup{E_n, \ldots, E_1}
  \end{eqnarray*}
\end{enumerate}
We would like to define $\variadic{R}$, the \agengen/ of $R_n$, such
that $(\variadic{R}\ c_n) = R_n$. We start by writing $R_{n+1}$ in terms of $R_n$:
\begin{eqnarray*}
R_{n+1} 
& = &
\lambda x_1 \cdots x_n x_{n+1} w.(w\ x_1 \cdots x_{n+1}) \\
& = & 
\lambda x_1 \cdots x_n x_{n+1} w.(R_n\ x_1 \cdots x_n\ (w\ x_{n+1})) \\
& = &
\lambda x_1 \cdots x_n x_{n+1} w.
(\comb B\ (R_n\ x_1 \cdots x_n)\ w\ x_{n+1}) \\
& = &
\lambda x_1 \cdots x_n x_{n+1} w.
(\comb C\ (\comb B\ (R_n\ x_1 \cdots x_n))\ x_{n+1}\ w) \\
& \eeta &
\lambda x_1 \cdots x_n.
(\comb C\ (\comb B\ (R_n\ x_1 \cdots x_n))) \\
& = &
\lambda x_1 \cdots x_n.
(\comb C\ (\comb B_n\ \comb B\ R_n\ x_1 \cdots x_n)) \\
& = &
\lambda x_1 \cdots x_n.
(\comb B_n\ \comb C\ (\comb B_n\ \comb B\ R_n)\ x_1 \cdots x_n) \\
& \eeta &
(\comb B_n\ \comb C\ (\comb B_n\ \comb B\ R_n))
\end{eqnarray*}
The \lt/ $f$ that takes a Church numeral $c_n$, and maps $\Maps {R_n} {(f\ c_n)} {R_{n+1}}$, is given by:
\begin{eqnarray*}
f & \equiv & 
\lambda n r.
(\vB\ n\ \comb C\ (\vB\ n\ \comb B\ r))
\end{eqnarray*}
The \lt/ $g$ such that $\Maps {\Ntup{c_n, R_n}} g {\Ntup{c_{n+1}, R_{n+1}}}$ is given by:
\begin{eqnarray*}
g & \equiv &
\lambda p . \Ntup{(\Succ\ (\pi_1^2\ p)), (f\ (\pi_1^2\ p)\ (\pi_2^2\ p))}
\end{eqnarray*}
Notice that $R_0 = \comb I$, so we can obtain $R_n$ by applying the $n$-th composition of $g$ to $\Ntup{c_0, \comb I}$. We define $\variadic R$ as follows:
\begin{eqnarray*}
\variadic{R} & \equiv & \lambda n.(\pi_2^2\ (n\ g\ \Ntup{c_0, \comb I}))
\end{eqnarray*}
Note that $(\variadic R\ c_n) = R_n$. 

Below are examples of two slightly different ways of using $\variadic{R}$:
\begin{eqnarray*}
(\variadic{R}\ c_3\ E_1\ E_2\ E_3) 
& = & (R_3\ E_1\ E_2\ E_3) \\
& = & \Ntup{E_3, E_2, E_1} \\
(\Ntup{E_1, E_2, E_3, E_4}\ (\variadic{R}\ c_4)) 
& = & (\Ntup{E_1, E_2, E_3, E_4}\ R_4) \\
& = & \Ntup{E_4, E_3, E_2, E_1}
\end{eqnarray*}

\subsection{Mapping}

We would like to define the combinator $\vMap$, such that: 
\begin{eqnarray*}
(\vMap\ c_n\ f\ \Ntup{x_1, \ldots, x_n}) & = &
\Ntup{(f\ x_1), \ldots (f\ x_n)}
\end{eqnarray*}
Let:
\begin{eqnarray*}
Q_n & \equiv & \lambda x_1 \cdots x_n z . (z\ (f\ x_1) \cdots (f\ x_n)) \\
\mathrm{Map}_n & \equiv & \lambda f v.(v\ Q_n)
\end{eqnarray*}
We define $Q_{n+1}$ in terms of $Q_n$:
\begin{eqnarray*}
Q_{n+1} & = & \lambda \vs x1n w z.(Q_n\ x_1\ \cdots x_n\ z\ (f\ w)) \\
& = & \lambda \vs x1n w z.(\comb C\ (Q_n\ \vs x1n)\ (f\ w)\ z) \\
& \eeta & \lambda \vs x1n w. (\comb C\ (Q_n\ \vs x1n)\ (f\ w)) \\
& = & \lambda \vs x1n w.(\comb B\ (\comb C\ (Q_n\ \vs x1n))\ f\ w) \\
& \eeta & \lambda \vs x1n.(\comb B\ (\comb C\ (Q_n\ \vs x1n))\ f) \\
& = & \lambda \vs x1n.(\comb C\ \comb B\ f\ (\comb C\ (Q_n\ \vs x1n))) \\
& = & \lambda \vs x1n.(\comb B\ (\comb C\ \comb B\ f)\ \comb C\ (Q_n\ \vs x1n)) \\
& = & \lambda \vs x1n.(\comb B_n\ (\comb B\ (\comb C\ \comb B\ f)\ \comb C)\ Q_n\ \vs x1n) \\
& \eeta & (\comb B_n\ (\comb B\ (\comb C\ \comb B\ f)\ \comb C)\ Q_n)
\end{eqnarray*}
Using this relation, we define the \lt/ $g$ to map $\Maps {\Ntup{c_n, Q_n}} g {\Ntup{c_{n+1}, Q_{n+1}}}$ for all $n$:
\begin{eqnarray*}
g & \equiv & 
\lambda p.\langle
\ver{(\Succ\ (\pi_1^2\ p)), }
    {(\vB\ (\pi_1^2\ p)\ 
           (\comb B\ (\comb C\ \comb B\ f)\ \comb C)\ 
           (\pi_2^2\ p))\rangle}@
\end{eqnarray*}
Notice that $Q_0 = \comb I$, so we can obtain $Q_n$ by applying the $n$-th composition of $g$ to $\Ntup{c_0, \comb I}$. We define $\variadic Q$ as follows:
\begin{eqnarray*}
\variadic Q & \equiv & 
\lambda n.(\pi_2^2\ (n\ g\ \Ntup{c_0, \comb I}))
\end{eqnarray*}

We now define $\vMap$ as follows:
\begin{eqnarray*}
\vMap & \equiv & \lambda n f v.(v\ (\variadic Q\ n))
\end{eqnarray*}
This definition satisfies the requirement that $(\vMap\ c_n) = \mathrm{Map}_n$.

\subsection{\Agen/, multiple fixed-point combinators}
\label{ssec:mfpcs}

{\sloppy
By now we have the tools nee- ded to construct \agen/, multiple 
fixed-point combinators in the \lc/.  Fixed-point combinators are used to solve fixed-point equations, resulting in a single solution that is the \textit{least} in a lattice-theoretic sense. When moving to $n$ multiple fixed-point equations, multiple fixed-point combinators are needed to solve the system, giving a set of $n$ solutions, that once again, are the least in the above-mentioned lattice-theoretic sense. 
}

A set of $n$ multiple fixed-point combinators are \lt/s $F_1^n,
\ldots, F_n^n$, such that for any $n$ \lt/s $x_1, \ldots, x_n$, and $k
= 1, \ldots, n$ we have:
\begin{eqnarray*}
  (F_j^n\ x_1 \cdots x_n) & = &
  (x_j\ (F_1^n\ x_1 \cdots x_n) \cdots (F_n^n\ x_1 \cdots x_n))
\end{eqnarray*}

Brevity is one motivation for the construction of an \agen/ fixed-point combinator. Using ordinary multiple fixed-point combinators, $n$ combinators are needed for \textit{any} choice of $n$, which means that if we wish to solve several such systems of equations, we need a great many number of multiple fixed-point combinators. In contrast, an \agen/ fixed-point combinator can be used to find \textit{any} multiple fixed-point in a system of \textit{any} size: It takes  as arguments two Church numerals $c_n, c_k$, which specify the size of the system, and the specific multiple fixed-point, and returns the specific multiple fixed-point combinator of interest.

Other reasons for using an \agen/ fixed-point combinator have to do with the size of the multiple fixed-point combinators and their correctness: The size of the $n$-ary extensions of Curry's and Turing's historical fixed-point combinators is \textit{quadratic} to the number of equations, or $O(n^2)$. Specifying such large terms, be in on paper, in \LaTeX, or in a computerized reduction system is unwieldy and prone to errors. An \agen/ fixed-point combinator is surprisingly compact, because the size of the system is specified as an argument. 

\subsubsection[An arity-generic generalization of Curry's
  combinator for multiple fixed points]{An \agengen/ of Curry's
  fixed-point combinator for multiple fixed points}
\label{sssec:curryfpc}

Recall Curry's single fixed-point combinator:
\begin{eqnarray*}
\Ycurry & \equiv & 
\lambda f.(\ver{(\lambda x.(f\ (x\ x)))}
               {(\lambda x.(f\ (x\ x))))}@
\end{eqnarray*}
Generalizing Curry's single fixed-point combinator to $n$ multiple fixed-point equations yields a sequence $\Set{\Phi_k^n}_{k = 1}^n$ of $n$ multiple fixed-point combinators, where $\Phi_k^n$ is defined as follows:
\begin{eqnarray*}
\Phi_k^n & \equiv &
\lambda f_1 \cdots f_n . (%
\ver{(\lambda x_1 \cdots x_n . (f_k\ (x_1\ x_1 \cdots x_n) \cdots (x_n\ x_1 \cdots x_n)))}
    {(\lambda x_1 \cdots x_n . (f_1\ (x_1\ x_1 \cdots x_n) \cdots (x_n\ x_1 \cdots x_n)))}
    {\hskip10em\vdots}
    {(\lambda x_1 \cdots x_n . (f_n\ (x_1\ x_1 \cdots x_n) \cdots (x_n\ x_1 \cdots x_n))))}@
\end{eqnarray*}
Given the system of fixed-point equations $\Set{(F_k\ x_1 \cdots x_n) = x_k}_{k = 1}^n$, the $k$-th multiple fixed-point is given by $(\Phi_k^n\ F_1\ \cdots F_n)$. 

Our inductive definition (on the syntax of \lc/) is sufficiently precise and well-defined that we can construct, for any given $n \in \mathbb{N}$, a set of multiple fixed-point combinators. But if $n$ is a variable, rather than a constant, then this will not do. 

Let $v_x = \Ntup{x_1, \ldots, x_n}$. Starting with the inner common sub-expression $\Ntup{(x_k\ x_1 \cdots
  x_n)}_{k=1}^n$, we note that: 
\begin{eqnarray*}
\Ntup{(x_k\ x_1 \cdots x_n)}_{k=1}^n & = & 
(\variadic{\mathrm{Map}}\ c_n\ (\lambda x_k.(x_k\ v_x))\ v_x) 
\end{eqnarray*}
The \agen/ fixed-point combinator $\vPhi$ takes $c_n, c_k$, and
returns $\Phi_k^n$, which is the fixed-point combinator that takes $n$
generating functions, and returns the $k$-th of $n$ multiple
fixed-points:
\begin{eqnarray*}
(\vPhi\ c_k\ c_n) & = &
\lambda f_1 \cdots f_n . (
\ver{(\lambda v_f.(
       \ver{(\lambda w.(\vpi\ c_k\ c_n\ w\ w))}
           {(\vMap\ \ver{c_n}
                        {(\lambda f_j v_x.(\vMap\ \ver{c_n}
                                                      {(\lambda x_k.(x_k\ v_x))}
                                                      {v_x\ f_j))}@}
                        {v_f)))}@}@}
    {(\vNtupMaker\ c_n\ f_1 \cdots f_n)}@ \\
& = &
(\vB\ \ver{c_n}
          {(\lambda v_f.(
              \ver{(\lambda w.(\vpi\ c_k\ c_n\ w\ w))}
                  {(\vMap\ \ver{c_n}
                               {(\lambda f_j v_x.(\vMap\ \ver{c_n}
                                                             {(\lambda x_k.(x_k\ v_x))}
                                                             {v_x\ f_j))}@}
                               {v_f)))}@}@}
          {(\vNtupMaker\ c_n))}@
\end{eqnarray*}
Abstracting the variables $k, n$ over $c_k, c_n$ respectively, we define the \agen/ extension of Curry's multiple fixed-point combinator:
\begin{eqnarray*}
\vPhi & \equiv & 
\lambda k n.
(\vB\ \ver{n}
          {(\lambda v_f.(
              \ver{(\lambda w.(\vpi\ k\ n\ w\ w))}
                  {(\vMap\ \ver{n}
                               {(\lambda f_j v_x.(\vMap\ \ver{n}
                                                             {(\lambda x_k.(x_k\ v_x))}
                                                             {v_x\ f_j))}@}
                               {v_f)))}@}@}
          {(\vNtupMaker\ n))}@
\end{eqnarray*}
This definition satisfies the requirement that $(\vPhi\ c_k\ c_n) = \Phi_k^n$.

\subsubsection[An arity-generic generalization of Turing's
  combinator for multiple fixed points]{An \agengen/ of Turing's
  fixed-point combinator for multiple fixed points}
\label{sssec:turingfpc}

Recall Turing's single fixed-point combinator:
\begin{eqnarray*}
\Yturing & \equiv & 
(\ver{(\lambda x f.(f\ (x\ x\ f)))}
       {(\lambda x f.(f\ (x\ x\ f))))}@
\end{eqnarray*}
Generalizing Turing's single fixed-point combinator to $n$ multiple
fixed-point equations yields a sequence $\Set{\Psi_k^n}_{k = 1}^n$ of
$n$ multiple fixed-point combinators, where $\Psi_k^n$ is defined as
follows:
\begin{eqnarray*}
\Psi_k^n & \equiv &
(\ver%
  {(\lambda x_1 \cdots x_n f_1 \cdots f_n . 
      (f_k\ (x_1\ x_1 \cdots x_n\ f_1 \cdots f_n) \cdots 
             (x_n\ x_1 \cdots x_n\ f_1 \cdots f_n)))}
{(\lambda x_1 \cdots x_n f_1 \cdots f_n . 
      (f_1\ (x_1\ x_1 \cdots x_n\ f_1 \cdots f_n) \cdots 
             (x_n\ x_1 \cdots x_n\ f_1 \cdots f_n)))}
  {\cdots}
{(\lambda x_1 \cdots x_n f_1 \cdots f_n . 
      (f_n\ (x_1\ x_1 \cdots x_n\ f_1 \cdots f_n) \cdots 
             (x_n\ x_1 \cdots x_n\ f_1 \cdots f_n))))}@
\end{eqnarray*}
Our construction follows similar lines as with the $n$-ary
generalization of $\Ycurry$. For a given $n$, the
ordered $n$-tuples $v_x, v_f$ are defined as follows:
\begin{eqnarray*}
v_x & \equiv & \Ntup{x_1, \ldots, x_n} \\
v_f & \equiv & \Ntup{f_1, \ldots, f_n} 
\end{eqnarray*}
respectively.

As before, we begin by encoding a common sub-expression
$\Ntup{(x_k\ x_1 \cdots x_n\ f_1 \cdots f_n)}_{k = 1}^{n}$, as
follows:
\begin{eqnarray*}
\Ntup{(x_k\ x_1 \cdots x_n\ f_1 \cdots f_n)}_{k = 1}^{n}
& = & (\vMap\ c_n\ (\lambda x_k . (x_k\ v_x\ v_f))\ v_x)
\end{eqnarray*}

The \agengen/ of Turing's multiple fixed-point combinator is given by:
\begin{eqnarray*}
(\vPsi\ c_k ~ c_n) & = &
\lambda f_1 \cdots f_n.(
\ver{(\lambda v_f.(
       \ver{(\lambda w.(\vpi\ c_k\ c_n\ w\ w\ v_f))}
           {(\vMap\ 
\ver{c_n}
    {(\lambda j v_x v_f.(\vMap\ 
       \ver{c_n}
           {(\lambda x_k.(x_k\ v_x\ v_f))}
           {v_x}
           {(\vpi\ (\Succ\ j)\ c_n\ v_f)))}@}
    {(\iota\ c_n)}@}@}
    {(\vNtupMaker\ c_n\ f_1 \cdots f_n))}@ \\
& = &
(\vB\ \ver{c_n}
{(\lambda v_f.(
       \ver{(\lambda w.(\vpi\ c_k\ c_n\ w\ w\ v_f))}
           {(\vMap\ 
\ver{c_n}
    {(\lambda j v_x v_f.(\vMap\ 
       \ver{c_n}
           {(\lambda x_k.(x_k\ v_x\ v_f))}
           {v_x}
           {(\vpi\ (\Succ\ j)\ c_n\ v_f)))}@}
    {(\iota\ c_n)}@}@}
          {(\vNtupMaker\ c_n))}@
\end{eqnarray*}
We define $\vPsi$ by abstracting $c_k, c_n$ over the above, to get:
\begin{eqnarray*}
\vPsi & \equiv &
\lambda k n.
(\vB\ \ver{n}
{(\lambda v_f.(
       \ver{(\lambda w.(\vpi\ k\ n\ w\ w\ v_f))}
           {(\vMap\ 
\ver{n}
    {(\lambda j v_x v_f.(\vMap\ 
       \ver{n}
           {(\lambda x_k.(x_k\ v_x\ v_f))}
           {v_x}
           {(\vpi\ (\Succ\ j)\ n\ v_f)))}@}
    {(\iota\ n)}@}@}
          {(\vNtupMaker\ n))}@
\end{eqnarray*}
This definition satisfies the requirement that $(\vPsi\ c_k\ c_n) = \Psi_k^n$. 

\subsubsection{An \agengen/ of \boehm/'s construction}

In Sections \ref{sssec:curryfpc} and \ref{sssec:turingfpc} we
introduced $n$-ary generalizations of Curry's and Turing's \fpc/ for
solving systems of multiple fixed-point equations. The goal of this
section is to show that these generalizations are, in a precise sense, natural,
and obey a well-known relation that holds between the two original,
single \fpc/s.

In his textbook \textit{The Lambda Calculus: Its Syntax and
  Semantics}~\cite[page~143]{Barendregt1984}, Barendregt mentions, in
the proof of Proposition~6.5.5, a result due to \boehm/, that relates Curry's and Turing's \fpc/s:

Let $M \equiv \lambda \phi x.(x\ (\phi\ x)) = (\comb S\ \comb I)$. We
have:
\begin{eqnarray*}
  (\Ycurry\ M) & \RReducesTo & \Yturing
\end{eqnarray*}

To understand whence this \lt/ $M$ comes, consider the definition of a \fpc/: A term $\Phi$, such that for all $x$, $(\Phi\ x)$ is a fixed point of $x$, and so we have:
\begin{eqnarray*}
(\Phi\ x) & = & (x\ (\Phi\ x))
\end{eqnarray*}
Abstracting over $x$, we get a recursive definition for $\Phi$, that can be rewritten as a fixed-point equation:
\begin{eqnarray*}
\Phi & = & \lambda x . (x\ (\Phi\ x)) \\
& = & ((\lambda \phi x . (x\ (\phi\ x)))\ \Phi) \\
& = & (M\ \Phi)
\end{eqnarray*}
We can solve this fixed-point equation using \textit{any} \fpc/. If $\Phi$ is a \fpc/, then $(\Phi\ M)$ is also a \fpc/. After we prove these to be distinct in the $\beta\eta$ sense, we can define an infinite chain of distinct \fpc/s. Furthermore, $M$ relates $\Ycurry$ and $\Yturing$ in an interesting way: $(\Ycurry\ M) \RReducesTo \Yturing$, which is a stronger relation than $=$. 

For the purpose of this work, we consider $n$-ary generalizations of $\Ycurry$ and $\Yturing$ to be natural if they satisfy a corresponding $n$-ary generalization of the above relation.

We now define $n$-ary generalizations of the above term $M$. If $\vsc{\Theta}1n$ are a set of $n$ multiple \fpc/s, then for any $\vsc x1n$ and $k = 1, \ldots, n$, it satisfies:
\begin{eqnarray*}
(\Theta_k^n\ \vs x1n) & = & (x_k\ (\Theta_1^n\ \vs x1n) \cdots (\Theta_n^n\ \vs x1n)) \\
& = & ( \ver{(\lambda \vs{\phi}1n\vs x1n.(x_k\ (\phi_1\ \vs x1n) \cdots (\phi_n\ \vs x1n)))}
                   {\vs{\Theta}1n)}@ \\
& = & (M_k^n\ \vs{\Theta}1n)
\end{eqnarray*}
where $M_k^n \equiv \lambda \vs{\phi}1n\vs x1n.(x_k\ (\phi_1\ \vs x1n)
\cdots (\phi_n\ \vs x1n))$.

The $n$-ary generalizations of $\Ycurry, \Yturing$ are given by $\Phi_k^n, \Psi_k^n$, respectively, for all $k = 1, \ldots, n$.

\begin{prop}\label{prop:mnk} For any $n > 0$ and each $k = 1, \ldots n$, we have $(\Phi_k^n\ M_1^n \cdots M_n^n) \RReducesTo \Psi_k^n$. 
\end{prop}

\proof 
\begin{eqnarray*}
\rlap{$(\Phi_k^n\ M_1^n \cdots M_n^n)$} \\& \RReducesTo &
( \ver{(\lambda \vs z1n.(M_k^n\ (z_1\ \vs z1n) \cdots (z_n\ \vs z1n)))}
        {(\lambda \vs z1n.(M_1^n\ (z_1\ \vs z1n) \cdots (z_n\ \vs z1n)))}
        {\cdots}
        {(\lambda \vs z1n.(M_n^n\ (z_1\ \vs z1n) \cdots (z_n\ \vs z1n))))}@ \\
& \RReducesTo &
( \ver{(\lambda \vs z1n \vs x1n . (x_k\ (z_1\ \vs z1n\ \vs x1n) \cdots (z_n\ \vs z1n\ \vs x1n)))}
        {(\lambda \vs z1n \vs x1n . (x_1\ (z_1\ \vs z1n\ \vs x1n) \cdots (z_n\ \vs z1n\ \vs x1n)))}
        {\cdots}
        {(\lambda \vs z1n \vs x1n . (x_n\ (z_1\ \vs z1n\ \vs x1n) \cdots (z_n\ \vs z1n\ \vs x1n))))}@ \\
& \equiv & \Psi_k^n\rlap{\hbox to 357 pt{\hfill\qEd}}
\end{eqnarray*}
We would like to define the combinator $\variadic{M}$, which is the \agengen/ of the $M_k^n$, such that:
\begin{eqnarray*}
(\variadic{M}\ c_k\ c_n) & = & M_k^n
\end{eqnarray*}

We start with $M_k^n$:
{
\def\tmp#1#2{%
(\vS\ \ver{c_n}
          {#1}
          {#2)}@
}
\begin{eqnarray*}
\def\tmp#1#2{%
(\vS\ \ver{c_n}
          {#1}
          {#2)}@
}
M_k^n & \equiv & 
\lambda \vs{\phi}1n\vs x1n .
(x_k\ (\phi_1\ \vs x1n) \cdots (\phi_n\ \vs x1n)) \\
& = & 
\lambda \vs{\phi}1n\vs x1n .
(\vsig\ c_k\ c_n\ \vs x1n\ (\phi_1\ \vs x1n) \cdots (\phi_n\ \vs x1n)) \\
& = &
\lambda \vs{\phi}1n.
\tmp{(\lambda \vs x1n.(\vsig\ c_k\ c_n\ \vs x1n\ (\phi_1\ \vs x1n) \cdots (\phi_{n-1}\ \vs x1n)))}{\phi_{n}} \\
& = &
\lambda \vs{\phi}1n.
\tmp{\tmp{(\lambda \vs x1n.(\vsig\ c_k\ c_n\ \vs x1n\ (\phi_1\ \vs x1n) \cdots (\phi_{n-2}\ \vs x1n)))}{\phi_{n-1}}}{\phi_{n}} \\
& = &
\lambda \vs{\phi}1n.
\underbrace{(\vS\ c_n\ (\vS\ c_n\ (\cdots(\vS\ c_n}_{\mbox{$n$ times}}\ (\vsig\ c_k\ c_n)\ \underbrace{\phi_1)\cdots)\ \phi_{n-1})\ \phi_n)}_{\mbox{$n$ times}}
\end{eqnarray*}
}%
We generate such a repeated application by repeatedly applying the
function $f$, defined so that 
$\Maps {\Ntup{M_r, c_r}} f {\Ntup{M_{r+1}, c_{r+1}}}$. 
Assuming the variable $n$, which stands for the Church numeral $c_n$ in the previous expression, and which occurs free in $f$, we define $f$ as follows:
\begin{eqnarray*}
f & = &
\lambda p.\langle 
\ver{(\vB\ (\pi_2^2\ p)\ (\vS\ n)\ (\pi_1^2\ p)),}
    {(\Succ\ (\pi_2^2\ p))\rangle}@
\end{eqnarray*}
We can now use $f$ to define $M_k^n$:
\begin{eqnarray*}
M_k^n & = &
(\pi_1^2\ (c_n\ f\ \Ntup{(\vsig\ c_k\ c_n), c_0}))
\end{eqnarray*}
We now define $\variadic{M}$ by abstracting $c_k, c_n$ over the parameterized expression, to get:
\begin{eqnarray*}
\variadic{M} & \equiv &
\lambda k n.(\pi_1^2\ (n\ f\ \Ntup{(\vsig\ k\ n), c_0})) \\
& \equiv & 
\lambda k n.
(\pi_1^2\ (n\ 
\ver{(\lambda p.\langle 
        \ver{(\vB\ (\pi_2^2\ p)\ (\vS\ n)\ (\pi_1^2\ p)),}
            {(\Succ\ (\pi_2^2\ p))\rangle)}@}
    {\Ntup{(\vsig\ k\ n), c_0}))}@
\end{eqnarray*}
This definition satisfies the requirement that $(\variadic{M}\ c_k\ c_n) = M_k^n$. Combined with Proposition~\ref{prop:mnk}, it follows that for $n \geq 1$ and for each $k = 1, \ldots, n$, we have: $(\vPhi\ c_k\ c_n\ (\variadic M\ c_k\ c_n)) = (\vPsi\ c_k\ c_n)$. The stronger $\RReducesTo$ property does not hold when working with encodings, which are by definition, $\beta$-equivalent. Finally, just as $M$ was used to construct a chain of infinitely-many different \fpc/s, so can $\variadic M$ be used to construct a chain of infinitely-many \agen/ \fpc/s: If $\Phi_1^n, \ldots, \Phi_n^n$ are $n$ multiple \fpc/s, then so are
{
  \def\B#1{%
    \ver{(\variadic M\ c_1\ c_n\ \Phi_1^n \cdots \Phi_n^n),}
        {\cdots}
        {(\variadic M\ c_n\ c_n\ \Phi_1^n \cdots \Phi_n^n)#1}@
  }
  \begin{eqnarray*}
    \B{}
  \end{eqnarray*}
  and so are
  \begin{eqnarray*}
    \ver{(\variadic M\ c_1\ c_n\ \B{),}}
        {\cdots}
        {(\variadic M\ c_n\ c_n\ \B)}@
  \end{eqnarray*}
  etc.
}

\subsubsection{Summary and conclusion}

We defined $n$-ary ($\Phi_k^n, \Psi_k^n$) and \agen/ ($\vPhi, \vPsi$) generalizations of Curry's and Turing's \fpc/s, and showed that these generalizations maintain the $n$-ary and \agengen/s of the relationship originally discovered by \boehm/. The significance of \agen/ \fpc/s is that they are \textit{single terms} that parameterize over the number of fixed-point equations and the index of a fixed point, so they can be used to find any fixed point of any number of fixed-point equations: They can be used interchangeably to define mutually-recursive procedures, mutually-recursive data structures, etc. 

For example, if $E, O$ are the \textit{even} and \textit{odd}
generating functions given by:
\begin{eqnarray*}
E & \equiv & \lambda eon.(\Zero\ n\ \comb{True}\ (o\ (\Pred\ n))) \\
O & \equiv & \lambda eon.(\Zero\ n\ \comb{False}\ (e\ (\Pred\ n)))
\end{eqnarray*}
Then we can use Curry's \agen/ \fpc/ to define the \lt/s that compute the \textit{even} and \textit{odd} functions on Church numerals as follows:
\begin{eqnarray*}
\mathrm{IsEven?} & \equiv & (\vPhi\ c_1\ c_2\ E\ O) \\
\mathrm{IsOdd?} & \equiv & (\vPhi\ c_2\ c_2\ E\ O)
\end{eqnarray*}
Alternatively, we can use Turing's \agen/ \fpc/ to do the same:
\begin{eqnarray*}
\mathrm{IsEven?}' & \equiv & (\vPsi\ c_1\ c_2\ E\ O) \\
\mathrm{IsOdd?}' & \equiv & (\vPsi\ c_2\ c_2\ E\ O)
\end{eqnarray*}

It might seem intuitive that in order to generate $n$ multiple fixed
points, we would need $n$ generating expressions, and this intuition
is responsible for the $O(n^2)$ size of the $n$-ary extensions of
Curry's and Turing's \fpc/s. A more compact approach, however, is to
pass along a single aggregation of the $n$ fixed points, which can be
done using a single generator function that is applied to itself. This
approach was taken by Kiselyov~\cite{Kiselyov:02} in his construction
of a variadic, multiple \fpc/ in Scheme:
\begin{minipage}[t]{\linewidth}
\begin{lstlisting}[language=scheme]
(define Y*
  (lambda s
    ((lambda (u) (u u))
     (lambda (p)
       (map (lambda (si) 
              (lambda x 
                (apply (apply si (p p)) x))) 
            s)))))
\end{lstlisting}
\end{minipage}
A corresponding \agen/ version can be encoded in the \lc/ in two ways. First, to emphasize the brevity of this construction, we can write:
{\def\LxDxx{(\lambda u.(u\ u))}
\begin{eqnarray*}
  \Yoleg & = &
  \lambda n v_s.
  (\ver{\LxDxx}
       {(\lambda p.(\vMap\ n\ (\lambda s_i v_x.(p\ p\ s_i\ v_x))\ v_s)))}@ \\
\end{eqnarray*}
Note that since the $\mathrm{Apply}$ combinator reverses its two
arguments, we can avoid it altogether by reversing its two arguments
\textit{in situ}, essentially \textit{inlining} the $\mathrm{Apply}$
combinator. Then for any $n \in \mathbb{N}$, let $f_1, \ldots f_n \in
\Lambda$ be some \le/s, and let $\Phi_1^n, \ldots, \Phi_n^n$ be a set
of $n$ multiple \fpc/s, $\Yoleg$ satisfies:
\begin{eqnarray*}
  (\Yoleg\ c_n\ \Ntup{f_1, \ldots f_n}) & = &
  \Ntup{(\Phi_1^n\ f_1 \cdots f_n), \ldots, (\Phi_n^n\ f_1 \cdots f_n)}
\end{eqnarray*}
But to be consistent with how we defined and used other \agen/ terms, we should rather define a Curried variant $\YolegVar$:
\def\Aii{(\lambda p s_i v_x.(p\ p\ s_i\ v_x))}
\def\Ai{(\comb C\ (\comb B\ \ver{(\vMap\ n)}
                                {\Aii))}@}
\begin{eqnarray*}
  \YolegVar & = &
  \lambda n.
  (\vB\ n\ \ver{\LxDxx}
               {(\vB\ n\ \ver{\Ai}
                             {(\vNtupMaker\ n)))}@}@
\end{eqnarray*}
This variant takes a Church numeral, followed by $n$ \le/s, and returns the $n$-tuple of their multiple fixed points:
\begin{eqnarray*}
  (\YolegVar\ c_n\ f_1 \cdots f_n) & = &
  \Ntup{(\Phi_1^n\ f_1 \cdots f_n), \ldots, (\Phi_n^n\ f_1 \cdots f_n)}
\end{eqnarray*}
}

So it seems that the shortest known multiple fixed-point combinator in
Scheme translates to a very short multiple fixed-point combinator in
the \lc/, perhaps the shortest known as well.

\subsection{Derivation of the \AGEN/~One-Point Basis Maker}

In a previous \linebreak work~\cite{Goldberg2004}, we have shown that for any $n$
\lt/s $E_1, \ldots, E_n$, which need not even be combinators, it is
possible to define a single term $X$ that generates $E_1, \ldots,
E_n$. Such a term is known as a \textit{one-point basis}~\cite[Section~8.1]{Barendregt1984}.

It is straightforward to construct a \textit{dispatcher} \lt/  $D$, such that $(D~c_k) = E_k$, for all $k = 1, \ldots, n$. Let $X = \Ntup{M, c_0}$, where $M = \lambda mba.(\Zero~b~\Ntup{m, (\Succ~a)}~(D~b))$. Then, for any $k = 1, \ldots, n$, we have:
\begin{eqnarray*}
X(\underbrace{X \cdots X}_{k+1}) 
& = & \Ntup{M, c_0}(\underbrace{\Ntup{M, c_0} \cdots \Ntup{M, c_0}}_{k+1}) \\
& = & \Ntup{M, c_0}\Ntup{M, c_k} \\
& = & (D~c_k) \\
& = & E_k
\end{eqnarray*}
Notice that a different dispatcher is needed for each $n$, and for each $E_1, \ldots, E_n$. 

Using our \agen/ basis, we can abstract a Church numeral over our construction, and obtain an \agen/ one-point basis \textit{maker}. We define $M$ so as to use an \agen/ selector to dispatch over $n$ expressions:
\begin{eqnarray*}
M 
& \equiv &
\lambda mba.(\Zero\ b\  
(\lambda x.x\ m\ (\Succ\ a))\  
(\vsig\ b\ c_n\ x_1 \cdots x_n))
\end{eqnarray*}
We use $M$ to define the \AGEN/ basis maker $\variadic{\mathrm{MakeX}}$:
\begin{eqnarray*}
\variadic{\mathrm{MakeX}} & = &
\lambda n x_1 \cdots x_n . \Ntup{M, c_0} \\
& = &
\lambda n x_1 \cdots x_n z . (z\ M\ c_0) \\
& = &
\lambda n x_1 \cdots x_n z . (\comb I\ z\ M\ c_0) \\
& = &
\lambda n x_1 \cdots x_n z . (\comb C\ \comb I\ M\ z\ c_0) \\
& = &
\lambda n x_1 \cdots x_n z . 
(\comb C\ (\comb C\ \comb I\ M)\ c_0\ z) \\
& \eeta &
\lambda n x_1 \cdots x_n . 
(\comb C\ (\comb C\ \comb I\ M)\ c_0) \\
& = &
\lambda n x_1 \cdots x_n . 
(\comb C\ \comb C\ c_0\ (\comb C\ \comb I\ M)) \\
& = &
\lambda n x_1 \cdots x_n . 
(\alias{\comb B\ (\comb C\ \comb C\ c_0)\ (\comb C\ \comb I)}{A_1}\ M) \\
& = &
\lambda n x_1 \cdots x_n . 
(A_1\ (\lambda mba.(\Zero\ b\ 
 \ver{(\lambda x.(x\ m\ (\Succ\ a)))}
     {(\vsig\ b\ n\ x_1 \cdots x_n))))}@ \\
& = &
\lambda n x_1 \cdots x_n . 
(A_1\ (\lambda mba.(\vB\ n\ 
\ver{(\Zero\ b\ (\lambda x.(x\ m\ (\Succ\ a))))}
    {\hskip-3em\alias{\hskip3em(\vsig\ b\ n)\hskip8.5em}{A_2}}
    {x_1 \cdots x_n)))}@ \\
& = &
\lambda n x_1 \cdots x_n . 
(A_1\ (\lambda mba.(A_2\ x_1 \cdots x_n))) \\
& = &
\lambda n x_1 \cdots x_n . 
(A_1\ (\lambda mba.(\Ntup{x_1, \ldots, x_n}\ A_2))) \\
& = &
\lambda n x_1 \cdots x_n . 
(A_1\ (\alias{(\lambda vmba.(v\ A_2))}{A_3}\ \Ntup{x_1, \ldots, x_n}) \\
& = &
\lambda n x_1 \cdots x_n . 
(A_1\ (A_3\ (\vNtupMaker\ n\ x_1 \cdots x_n))) \\
& = &
\lambda n x_1 \cdots x_n . 
(A_1\ (\alias{\vB\ n\ A_3\ (\vNtupMaker\ n)}{A_4}\ x_1 \cdots x_n)) \\
& = &
\lambda n x_1 \cdots x_n . 
(\vB\ n\ A_1\ A_4\ x_1 \cdots x_n) \\
& \eeta &
\lambda n . (\vB\ n\ A_1\ A_4) \\
& = &
\lambda n . (\vB\ n\ 
\ver{A_1}
    {(\unalias{\vB\ n\ A_3\ (\vNtupMaker\ n)}{A_4}))}@ \\
& = &
\lambda n . (\vB\ n\ 
\ver{A_1}
    {(\vB\ n\ (\unalias{(\lambda vmba.(v\ A_2)}{A_3})\ (\vNtupMaker\ n)))}@ \\
& = &
\lambda n . (\vB\ n\ 
\ver{A_1}
    {(\vB\ n\ (
\ver{(\lambda vmba.(v\ (\vB\ n\ 
\ver{(\Zero\ b\ (\lambda x.(x\ m\ (\Succ\ a))))}
    {\hskip-3em\unalias{\hskip3em(\vsig\ b\ n))))\hskip8.5em}{A_2}}@}
    {(\vNtupMaker\ n))))}@}@ \\
& = &
\lambda n . (\vB\ n\ 
\ver{(\unalias{\comb B\ (\comb C\ \comb C\ c_0)\ (\comb C\ \comb I)}{A_1})}
    {(\vB\ n\ (
\ver{(\lambda vmba.(v\ (\vB\ n\ 
\ver{(\Zero\ b\ (\lambda x.(x\ m\ (\Succ\ a))))}
    {(\vsig\ b\ n))))}@}
    {(\vNtupMaker\ n))))}@}@
\end{eqnarray*}
We may now define $\variadic{\mathrm{MakeX}}$ as follows:
\begin{eqnarray*}
\variadic{\mathrm{MakeX}}
& \equiv &
\lambda n . (\vB\ n\ 
\ver{(\comb B\ (\comb C\ \comb C\ c_0)\ (\comb C\ \comb I))}
    {(\vB\ n\ (
\ver{(\lambda vmba.(v\ (\vB\ n\ 
\ver{(\Zero\ b\ (\lambda x.(x\ m\ (\Succ\ a))))}
    {(\vsig\ b\ n))))}@}
    {(\vNtupMaker\ n))))}@}@ \\
\end{eqnarray*}
We can use $\variadic{\mathrm{MakeX}}$ as follows. For any $n > 1$ and $E_1, \ldots, E_n \in \Lambda$, we can define $X$ as follows:
\begin{eqnarray*}
X & \equiv & (\variadic{\mathrm{MakeX}}\ c_n\ E_1 \cdots E_n)
\end{eqnarray*}
We now have:
\begin{eqnarray*}
(X\ (X\ X)) & = & E_1 \\
(X\ (X\ X\ X)) & = & E_2 \\
& \cdots & \\
(X\ \underbrace{(X \cdots X)}_{n+1}) & = & E_n \\
\end{eqnarray*}
Notice that we have made no assumptions about $E_1 \ldots E_n$, and in particular, have not required that they be combinators. Our one-point basis maker, $\variadic{\mathrm{MakeX}}$, provides an abstract mechanism for packaging \lt/s, in a way that they can later be ``unpacked''. 

\subsection{Summary and Conclusion}

We used our extended basis and \ba/ algorithm to encode useful \agen/ \lt/s of increasing complexity. We took the approach that working with sequences of expressions in an intuitive, modular and systematic way should resemble ``list processing'' known from \lisp/ and other functional programming languages. 

In the spirit of list processing, the first part of this section introduces \agen/ \lt/s for picking elements of sequences, constructing ordered $n$-tuples, applying \lt/s to the elements of a tuple, extending and reversing tuples, and constructing new ordered $n$-tuples by \textit{mapping} over existing tuples. All these \lt/s correspond to the basic machinery for list processing, e.g., in \lisp/. Once these were defined, we were ready to look at more complex \agen/ \lt/s.

Our detailed examples include \agen/ \fpc/s, and an \agen/ generator for one-point bases. 

We encoded \agengen/s of two historical \fpc/s by Curry and Turing. These \fpc/s maintain a relationship discovered by \boehm/, so it is natural to wonder whether this relationship is maintained in the \agengen/s of these \fpc/s, and we have shown this to be the case up to $\beta$-equivalence. 

We then encoded an \agen/ generator for one-point bases, so that any number of \lt/s can be ``compacted'' into a single expression from which they can be generated.

We tested all the \agen/ definitions in this work using a normal-order reducer for the \lc/, and have verified that they behave as expected on an array of examples.

\section{Related Work}

The expressive power of the \lc/ has fostered the advent of functional
languages.
For example, the Algorithmic Language Scheme~\cite{Sussman-Steele:75}
was developed as an interpreter for the \lc/, and offered programmatic
support for playing with $\lambda$-definability, from Church numerals
to a call-by-value version of Curry's fixed-point
combinator~\cite{Steele-Sussman:TR76-imperative}.
Since Scheme provides linguistic support for variadic functions, it has
become a sport to program call-by-value fixed-point operators for
variadic functions.
Queinnec presented the Scheme procedure \verb|NfixN2|, that is a
variadic, applicative-order multiple fixed-point
combinator~\cite[Pages~457--458]{Queinnec1996}.
The author presented one that directly extends Curry's fixed-point
combinator~\cite{Goldberg:HOSC05} and was a motivation for
Section~\ref{ssec:mfpcs}.

The original aim of the Combinatory-Logic program, as pursued by
\Schon/~\cite{Schoenfinkel1924}, was the elimination of bound
variables~\cite{CardoneHindley2009}. To this end, \Schon/ introduced
five constants, each with a conversion rule that described its
behavior. These constants are known today as $\ikbcs$ . While \Schon/
did not leave an explicit \textit{abstraction algorithm} for
translating terms with bound variables to equivalent terms without
bound variables~\cite[page~8]{Curry1958}, Cardon and Hindley claim it
extremely likely that he knew of such an
algorithm~\cite{CardoneHindley2009}.
 
As far as we have been able to verify, the first to have considered
the question of how to encode inductive and \agen/ \lt/s was Curry,
first in an extended Combinatory Logic framework~\cite{Curry1933},
where Curry first mentions such variables, and refers to them as
\textit{apparent variables}, and later, for Combinatory
Logic~\cite[Section~5E]{Curry1958}. We have not found this terminology
used elsewhere, and since the term \textit{\agen/} is much more
self-explanatory, we have chosen to stick with it.

Abdali, in his article \textit{An Abstraction Algorithm for Combinatory
  Logic}~\cite{Abdali1976}, presented a much simpler algorithm for
encoding inductive and \agen/ \lt/s. Abdali introduces the terms:
\begin{itemize}
\item $\mathscr{K}$, which is an \agengen/ of $\comb K$, and
identical to the $\vK$ combinator used throughout this article.
\item $\mathscr{I}$, which is an \agen/ selector, and is identical to
  the $\vsig$ combinator introduced in Section~\ref{ssec:sel}.
\item $\mathscr{B}$, which is a \textit{double} \agengen/
  of Curry's $\Phi = \lambda xyzu.(x\ (y\ u)\ (z\ u))$
  combinator~\cite{Curry1958}, generalized for two independent
  indices.
\end{itemize}
These combinators can augment any basis, and provide for a
straightforward encoding of \agen/ \lt/s. Abdali does not explain how he
came up with the double generalization of Curry's $\Phi$ combinator, or how
he encoded the definitions for $\mathscr{K}, \mathscr{I}, \mathscr{B}$
in terms of the basis he chose to use. \Agen/ expressions encoded using
$\mathscr{K}, \mathscr{I}, \mathscr{B}$, are not as concise as they
could be, because the $\mathscr{B}$ combinator introduces variables
even in when they are not needed in parts of an application, and in
such cases, a subsequent projection is needed to remove them.

Barendregt~\cite{Barendregt1984} seems to have considered this
question at least for some special cases, as in Exercises~8.5.13 and
8.5.20, the later of which he attributes to David
A. Turner.\footnote{Barendregt refers to Turner's article \textit{A
    New Implementation Technique for Applicative
    Languages}~\cite{Turner1979}, but as this article contains no
  mention of $n$-ary expressions and their encoding in the \lc/, it is
  plausible that he had really intended to refer to another article by
  Turner, also published in 1979: \textit{Another Algorithm for
    Bracket Abstraction}~\cite{Turner1979a}.}

\Schon/'s original $\ikbcs$ basis, coupled with Turner's \ba/
algorithm for that basis, offers several advantages in terms of
brevity of the resulting term, simplicity, intuitiveness and ease of
application of the algorithm. In the original \ba/ algorithm for
$\ikbcs$, the length of the encoded \lt/ is less than or equal to the
length of the original \lt/, because each application is replaced by a
combinator, and abstractions are either represented by a single
combinator, or are removed altogether through $\eta$-reduction. The
additional \agen/ combinators with which we extended the $\ikbcs$
basis maintain this conciseness, because a sequence of left-associated
applications to a sequence of variables is replaced by a single \agen/
combinator, and a sequence of Curried, nested \labs/s is either
removed via repeated $\eta$-expansions, or is replaced with by a
single \agen/ combinator. The extension of the basis and the
corresponding \ba/ algorithm to handle \agen/ \lt/s is straightforward
and intuitive.

\section{Discussion}

The ellipsis (`$\cdots$') and its typographical predecessor `\&c' (an
abbreviation for the Latin phrase \textit{et cetera}, meaning ``and
the rest'') have been used as meta-mathematical notation, to
abbreviate mathematical objects (numbers, expressions, formulae,
structures, etc.) for hundreds of years, going back to the 17th century
and possibly earlier. Such abbreviations permeate the writings of
Isaac Newton, John Wallis, Leonhard Euler, Carl Friedrich Gauss, and
up to the present. Despite its ubiquity, and perhaps as a paradoxical
tribute to this ubiquity, the ellipsis does not appear as an entry in
standard texts on the history of mathematical notation, even though
the authors of these texts make extensive use of ellipses in their
books~\cite{Cajori:1993,MazurJ:2014}. Neither is the ellipsis
discussed in the Kleene's classical text on
metamathematics~\cite{Kleene:1964}, nor does it even appear as an
entry in the list of symbols and notation at the end of the book, even
though Kleene makes extensive use of the ellipses both in the main
text as well as in the list of symbols and notation.

Discussions about the ellipsis and its meanings seem to concentrate in
computer literature: Roland Backhouse refers to the ellipsis as the
\textit{dotdotdot notation} in one of the more mathematical parts of
his book \textit{Program Construction: Calculating Implementations
  From Specifications}~\cite[Section~11.1]{Backhouse:2003}, and
suggests that they have many disadvantages, the most important being
that ``{\ldots}it puts a major burden on the reader, requiring them to
interpolate from a few example values to the general term in a bag of
values.'' Some of the examples of ellipses he cites can be rewritten
using summations, products, and the like. Others, however involve the
meta-language, e.g., functions that take $n$ arguments, where $n$ is a
meta-variable. Such examples of ellipses cannot be removed as easily.

The ellipsis also appears in some programming languages. In some
languages (C, C++, and Java) it is used to define variadic
procedures. In other languages (Ruby, Rust, and GNU extensions to C
and C++) it is used to define a range. In Scheme, the ellipsis is part
of the syntax for writing macros, which can be thought of as a
meta-language for Scheme. A formal treatment of ellipses in the macro
language for Scheme was done by Eugene Kohlbecker in his PhD
thesis~\cite{Kohlbecker:PhD}.

\Agen/ terms are somewhat reminiscent of \textit{variadic} procedures
in programming languages: The term \textit{variadic}, introduced by
Strachey~\cite{Strachey1967}, refers to the \textit{arity} of a
procedure, i.e., the number of arguments to which it can be applied. A
\textit{dyadic} procedure can be applied to two arguments. A
\textit{triadic} procedure can be applied to three arguments. A
\textit{variadic} procedure can be applied to any number of
arguments. Programming languages that provide a syntactic facility for
defining variadic procedures include C++ and \lisp/. The \lc/ has no
such syntactic facility, and so it is somewhat of a misnomer to speak
of \textit{variadic \lt/s}, since the number of arguments is an
\textit{explicit} parameter in our definitions, whereas in the
application of a variadic procedure to some arguments, the
\textit{number} of arguments is \textit{implicit} in an
implementation. Nevertheless, within the classical, untyped \lc/,
\agen/ \lt/s provide an expressivity that comes very close to having
variadic \lt/s.

Variadic procedures are not just about the procedure
\textit{interface}. When used in combination with \texttt{map} and
\texttt{apply}, they can provide a kind of generality that is
typically deferred to the meta-language or macro
system~\cite{Goldberg:HOSC05,Kohlbecker:PhD}. \Agen/ \ldef/ achieves
similar generality in the classical \lc/, with some notable
differences: Variadic procedures are applied to arbitrarily-many
arguments, and their parameter is bound to the list of the values of
these arguments. By contrast, \agen/ expressions take the number of
arguments, and return that many Curried $\lambda$-abstractions. In
this work, we used ordered $n$-tuples, rather than linked lists, as is
common in most functional programming languages, in what is perhaps
reminiscent of array programming languages. As a result of the choice
to use ordered $n$-tuples, the \textit{apply} operation became very
simple. It would be straightforward to choose to use linked lists
instead, at the cost of having to define \textit{apply} as a
\textit{left fold} operation.

In this work we show how to define, in the language of the \lc/,
expressions that contain meta-linguistic ellipses, the size of which
is indexed by a meta-variable. For such an indexed \lt/ $E_n$, our
goal was to find a term $\variadic E$ that takes $n$ as an explicit
parameter, and assuming it to be a Church numeral denoting the size of
the indexed expression, evaluates to $E_n$: $(\variadic E\ c_n) =
E_n$. We call $\variadic E$ an \agengen/ of $E$.

Of course, our choice of using Church numerals in this paper is based
on their ubiquity. In fact, any numeral system can be used, and we
have also constructed an \agen/ basis around Scott
numerals~\cite{Wadsworth80}.

Our approach has been to extend the basis $\Set{\ikbcs}$ with the
\agengen/s of $\comb K, \comb B, \comb C, \comb S$ combinators and to
extend Turner's \ba/ algorithm to handle abstractions of sequences of
variables over an expression. We then used this extended basis and
this extended \ba/ algorithm to encode \agen/ \lt/s. Our goal has not
been to remove all abstractions in \agen/ terms, but only those
abstractions that are over sequences of variables. Of course, it is
possible to remove all remaining abstractions, but our goal here has
been to define indexed expressions in the \lc/, without resorting to
meta-linguistic ellipses, for which the removal of all abstractions is
unnecessary.

In the first part of this work we presented a natural, \agengen/
to \Schon/'s $\Set{\ikbcs}$ basis for the set of combinators in the
\lkbec/, and extended Turner's \ba/ algorithm to make
use of the additional \agen/ combinators in the extended basis. The
extended algorithm retains the conciseness and simplicity of Turner's
original algorithm.

The second part of this work uses the \agen/ basis and the
corresponding \ba/ algorithm to develop tools for \agen/ \ldef/, and
incidentally de\-monstrates how the \agen/ basis can be used: We
introduced several \agen/ \lt/s that perform a wide variety of
computations on ordered $n$-tuples. These computations were inspired
by, and resemble to some extent, the facilities for \textit{list
  manipulation} that are native to the \lisp/ programming
language~\cite{Abelson85sussmanwith,Friedman1986,McCarthy1962}: Terms
that compute mappings, reversal, \agen/ fixed-point combinators,
\agen/ one-point bases, etc. Implementing in the \lc/ a functional
subset of the list processing capabilities of \lisp/ is a popular
exercise.

In his textbook on the \lc/, Barendregt states that there are two ways
to define ordered $n$-tuples: Inductively, using nested ordered pairs,
and another way, which Barendregt characterizes as being ``more
direct'', as $\Ntup{M_0, \ldots, M_n} = \lambda z.(z\ \vs
M0n)$~\cite[pages~133-134]{Barendregt1984}. Section~\ref{ssec:nTupleMaker}
shows how to make this more direct definition inductive.

In a previous work~\cite{Goldberg:HOSC05}, we derived an
applicative-order, variadic fixed-point combinator in Scheme. In that
work, we relied on Scheme's support for writing variadic procedures,
and consequently, on the primitive procedure \texttt{apply}, to apply
procedures to lists of their arguments. In the present work, we had
control over the representation of sequences, so we could encode an
\agen/ version of \texttt{apply}, as well as \agen/ \fpc/s, all within
the \lc/.

\section*{Acknowledgments}

The author is grateful to his anonymous reviewers and to his editor,
Neil D. Jones.  Thanks are also due to John Franco and Albert Meyer
for comments and questions about a previous work, and to Olivier Danvy
for his encouragement and suggestions.


\newpage

\begin{thebibliography}{10}

\bibitem{Abdali1976}
S.~Kamal Abdali.
\newblock An abstraction algorithm for combinatory logic.
\newblock {\em The Journal of Symbolic Logic}, 41(1):{222--224}, March 1976.

\bibitem{Abelson85sussmanwith}
Harold Abelson and Gerald Jay.
\newblock {\em Sussman with Julie Sussman. Structure and Interpretation of
  Computer Programs}.
\newblock MIT Press, 1985.

\bibitem{Backhouse:2003}
Roland Backhouse.
\newblock {\em Program Construction: Calculating Implementations from
  Specifications}.
\newblock John Wiley \& Sons, Inc., New York, NY, USA, 2003.

\bibitem{Barendregt1984}
Henk Barendregt.
\newblock {\em The Lambda Calculus: Its Syntax and Semantics}, volume 103 of
  {\em Studies in Logic and the Foundation of Mathematics}.
\newblock North-Holland, revised edition, 1984.

\bibitem{Cajori:1993}
Florian Cajori.
\newblock {\em A history of mathematical notations}.
\newblock Dover Publications, 1993.

\bibitem{CardoneHindley2009}
Felice Cardone and J.~Roger Hindley.
\newblock Lambda-calculus and combinators in the 20th century.
\newblock In Dov~M. Gabbay and John Woods, editors, {\em Logic from Russell to
  Church}, volume~5 of {\em Handbook of the History of Logic}, pages 723--817.
  North-Holland, 2009.

\bibitem{Church1941}
{A}lonzo {C}hurch.
\newblock {\em {T}he {C}alculi of {L}ambda-{C}onversion}.
\newblock {P}rinceton {U}niversity {P}ress, 1941.

\bibitem{Curry1933}
Haskell~B. Curry.
\newblock Apparent variables from the standpoint of combinatory logic.
\newblock {\em Annals of Mathematics}, 34(3):381--404, July 1933.

\bibitem{Curry1958}
{H}askell~{B}. {C}urry, {R}obert {F}eys, and {W}illiam {C}raig.
\newblock {\em {C}ombinatory {L}ogic}, volume~{I}.
\newblock {N}orth-{H}olland {P}ublishing {C}ompany, 1958.

\bibitem{Curry1972}
{H}askell~{B}. {C}urry, {J}.~{R}oger {H}indley, and {J}onathan~{P}. {S}eldin.
\newblock {\em {C}ombinatory {L}ogic}, volume~{I}{I}.
\newblock {N}orth-{H}olland {P}ublishing {C}ompany, 1972.

\bibitem{Friedman1986}
{D}aniel~{P}. {F}riedman and {M}atthias {F}elleisen.
\newblock {\em {T}he {L}ittle {L}{I}{S}{P}er}.
\newblock {S}cience {R}esearch {A}ssociates, {I}nc, 1986.

\bibitem{Goldberg2004}
Mayer Goldberg.
\newblock A construction of one-point bases in extended lambda calculi.
\newblock {\em Information Processing Letters}, 89(6):281 -- 286, 2004.

\bibitem{Goldberg:HOSC05}
Mayer Goldberg.
\newblock A variadic extension of {C}urry's fixed-point combinator.
\newblock {\em Higher-Order and Symbolic Computation}, 18(3/4):371--388, 2005.

\bibitem{Goldberg-lambda-calculus-tutorial}
Mayer Goldberg.
\newblock The {L}ambda {C}alculus: Outline of lectures., 2007-2011.
\newblock Department of Computer Science, Ben-Gurion University. Document URL:
  \verb|http://lambda.little-lisper.org/|.

\bibitem{Iverson1962}
{K}enneth~{E}. Iverson.
\newblock {\em {A} {P}rogramming {L}anguage}.
\newblock {J}ohn {W}iley \& {S}ons, {I}nc., 1962.

\bibitem{Kiselyov:02}
Oleg Kiselyov.
\newblock Simplest poly-variadic fix-point combinators for mutual recursion.
\newblock \verb|http://okmij.org/ftp/Computation/fixed-point-combinators.html|,
  2002.

\bibitem{Kleene1935a}
Stephen~C. Kleene.
\newblock A {T}heory of {P}ositive {I}ntegers in {F}ormal {L}ogic. {P}art {I}.
\newblock {\em American Journal of Mathematics}, 57(1):153--173, January 1935.

\bibitem{Kleene:1964}
Stephen~Cole Kleene.
\newblock {\em Introduction to Metamathematics}.
\newblock North-Holland Publishing, 1964.

\bibitem{Kohlbecker:PhD}
Eugene~E. Kohlbecker.
\newblock {\em Syntactic Extensions in the Programming Language Lisp}.
\newblock PhD thesis, Indiana University, Computer Science Department,
  Bloomington, Indiana, 1986.

\bibitem{MazurJ:2014}
Joseph Mazur.
\newblock {\em Enlightening Symbols: A Short History of Mathematical Notation
  and Its Hidden Powers}.
\newblock Princeton University Press, 2014.

\bibitem{McCarthy1962}
{J}ohn {M}c{C}arthy, {P}aul~{W}. {A}brahams, {D}aniel~{J}. {E}dwards,
  {T}imothy~{P}. {H}art, and {M}ichael~{I}. {L}evin.
\newblock {\em {L}{I}{S}{P} 1.5 {P}rogrammer's {M}anual}.
\newblock {M}{I}{T} {P}ress, {C}ambridge, {M}assachusetts, 1962.

\bibitem{Pakin1972}
Sandra Pakin.
\newblock {\em APL{\textbackslash}360 reference manual}.
\newblock Science Research Associates, Inc., 1972.

\bibitem{Queinnec1996}
{C}hristian {Q}ueinnec.
\newblock {\em {L}{I}{S}{P} {I}n {S}mall {P}ieces}.
\newblock {C}ambridge {U}niversity {P}ress, 1996.

\bibitem{Schoenfinkel1924}
Moses Sch\"onfinkel.
\newblock {\"Uber die Bausteine der mathematischen Logik}.
\newblock {\em {Mathematische Annalen}}, 92:{305--316}, 1924.
\newblock Translated by Stefan Bauer-Mengelberg as ``On the building blocks of
  mathematical logic'', in Jean van Heijenoort, 1967. \textit{A Source Book in
  Mathematical Logic}, 1879--1931. Harvard University Press. Pages~355--66.

\bibitem{Steele-Sussman:TR76-imperative}
Guy~L. {Steele Jr.} and Gerald~J. Sussman.
\newblock Lambda, the ultimate imperative.
\newblock AI Memo 353, Artificial Intelligence Laboratory, Massachusetts
  Institute of Technology, Cambridge, Massachusetts, March 1976.

\bibitem{Strachey1967}
Christopher Strachey.
\newblock Fundamental concepts in programming languages.
\newblock International Summer School in Computer Programming, Copenhagen,
  Denmark, August 1967.
\newblock Reprinted in Higher-Order and Symbolic Computation 13(1/2):11--49,
  2000.

\bibitem{Sussman-Steele:75}
Gerald~J. Sussman and Guy~L. {Steele Jr.}
\newblock Scheme: An interpreter for extended lambda calculus.
\newblock AI Memo 349, Artificial Intelligence Laboratory, Massachusetts
  Institute of Technology, Cambridge, Massachusetts, December 1975.
\newblock Reprinted in Higher-Order and Symbolic Computation 11(4):405--439,
  1998.

\bibitem{Turner1979a}
David~A. Turner.
\newblock Another algorithm for bracket abstraction.
\newblock {\em The Journal of Symbolic Logic}, 44(2):267--270, June 1979.

\bibitem{Turner1979}
David~A. Turner.
\newblock A new implementation technique for applicative languages.
\newblock {\em Software Practice and Experience}, 9(9):31--49, 1979.

\bibitem{Wadsworth80}
Christopher~P. Wadsworth.
\newblock Some unusual $\lambda$-calculus numeral systems.
\newblock In Jonathan~P. Seldin and J.~Roger Hindley, editors, {\em To H. B.
  Curry: Essays on Combinatory Logic, Lambda Calculus and Formalism}, pages
  215--230. Academic Press, London, 1980.

\bibitem{Weirich2010}
Stephanie Weirich and Chris Casinghino.
\newblock Arity-generic datatype-generic programming.
\newblock In {\em Proceedings of the 4th ACM SIGPLAN workshop on Programming
  languages meets program verification}, PLPV '10, pages 15--26, New York, NY,
  USA, 2010. ACM.

\end{thebibliography}
\end{document}